\documentclass[longbibliography,pre,amsfonts,notitlepage]{revtex4-1}
\usepackage{amsmath,amssymb,graphicx,tikz,natbib}
\usepackage{caption, xcolor}
\usepackage[colorlinks=true, citecolor=red, linkcolor=blue,urlcolor=blue ]{hyperref}

\AtBeginDocument{%
	\newwrite\bibnotes
	\def\bibnotesext{Notes.bib}
	\immediate\openout\bibnotes=\jobname\bibnotesext
	\immediate\write\bibnotes{@CONTROL{REVTEX41Control}}
	\immediate\write\bibnotes{@CONTROL{%
			apsrev41Control,author="08",editor="1",pages="1",title="0",year="1"}}
	\if@filesw
	\immediate\write\@auxout{\string\citation{apsrev41Control}}%
	\fi
}%

\def \bm#1{{\bf #1}}

\begin{document}
\def\v{\vspace{2cm}}
\title{Hydrodynamic synchronisation in strong confinement}

\author{Ivan Tanasijevi\'c}

\email[]{it279@cam.ac.uk}

\affiliation{Department of Applied Mathematics and Theoretical Physics,
	University of Cambridge, Cambridge CB3 0WA, UK}
\author{Eric Lauga}
\email[]{e.lauga@damtp.cam.ac.uk}
\affiliation{Department of Applied Mathematics and Theoretical Physics,
	University of Cambridge, Cambridge CB3 0WA, UK}

\begin{abstract}

Cellular appendages conferring motility, such as flagella and cilia, are known to  synchronise their periodic beats.  The origin of  synchronisation is a combination of  long-range hydrodynamic interactions with  physical mechanisms allowing the phases of these biological oscillators to evolve. Two of such mechanisms have been identified by previous work, the elastic compliance of the periodic orbit or oscillations driven by phase-dependent biological forcing, both of which can lead generically to stable phase locking. In order to help  uncover the{   physical mechanism for hydrodynamic synchronisation most essential overall} in biology, 
we theoretically investigate in this paper the effect of  strong  confinement on the effectiveness of   hydrodynamic synchronisation. 
Following past work, we use minimal models of cilia  where  appendages are modelled as rigid spheres forced to move along circular trajectories near a rigid surface. Strong confinement is modelled by  adding  a second nearby surface, parallel to the first one,  where the distance between the surfaces is much smaller than the typical distance between the cilia, which  results in a  qualitative change in the  nature of hydrodynamic interactions. We calculate  separately the impact of hydrodynamic confinement on  the synchronisation dynamics of  the elastic compliance and the force modulation   mechanisms and compare our results to the usual case with a single surface. Applying our results to the biologically relevant situation of nodal cilia, we show that  force modulation is a mechanism that leads to phase-locked states under  strong confinement that are very similar to those without confinement as a difference with the elastic compliance mechanism. Our results point therefore to the robustness of force modulation for synchronisation, an important feature for biological dynamics that therefore suggests it could be {the most essential physical mechanism overall in arrays of nodal cilia. We further examine the distinct  biologically-relevant situation of primary cilia and show in that case that the difference in robustness of the mechanisms is not as pronounced but still  favours   the force modulation.}

\end{abstract}

\date{\today}

\maketitle 

\newpage
\section{Introduction}

The world has long been fascinated by the effect of synchrony. 
Generally considered to be visually appealing, synchrony is  often exploited in  sports, such as rhythmic gymnastics or synchronised swimming, and in arts, such as dance and film. Scientists have also addressed synchronisation, starting with  the famous experiment of  Huygens in the 17th century who noticed  that   two pendulum clocks hung on a wall of a boat synchronise over time~\cite{huygens1899oeuvres}. The scientific interest in  synchrony naturally grew over time with more  examples observed in nature, such as the simultaneous flashing of fireflies~\cite{fireflies} that sparked the idea of self-organisation that seemed so ``contrary to all natural laws''~\cite{Laurent44}.  We later discovered that  synchronisation was essential for various functions of living organisms, for example the synchronisation between the human circadian clock and the daylight cycle~\cite{circadian} and it is also responsible for the establishment of the left-right asymmetry in early stage embryos~\cite{LRsymm}.

One aspect of synchronisation that recently received a lot of interest is that applied to cell motility, in particular in the interactions between swimming appendages. The flagella of many cells beat in synchrony~\cite{chlamy_sync,BERG1973} while   cilia  are known  to deform collectively as metachronal waves~\cite{paramecium,Blake_cilia,volvox,GueronMetachronal}, where neighbouring cilia beat with  fixed phase differences. 
On a scale of multiple microswimmers, hydrodynamic interactions can, for example, be responsible for the synchronisation of nearby   spermatozoa~\cite{sperm1, sperm2}. It was confirmed both experimentally~\cite{slips, kirsty,LightDriven,coloids} and via numerical simulations~\cite{GueronMetachronal,David} that hydrodynamic interactions between  appendages  are sufficient to induce synchronisation without the need for further   chemical coordination, an idea that was initially proposed by Taylor~\cite{taylor}. Note that, in some unicellular organisms,  synchronisation is mediated by intra-cellular  (basal) coupling~\cite{basal}. 
 
   Although hydrodynamic interactions often appear to be a necessary ingredient for synchronisation, they of course need to be combined with other physical mechanisms allowing the phase of oscillators to change with time~\cite{KimRigidHelix}. As a consequence,  additional physical  features of flagella and cilia  have to be identified, which motivated a lot of theoretical and experimental work~\cite{GueronFirst,kirstyReview,GolestanianRev,LironSim,Joanny}.{  Numerous theoretically studies  on this topic modelled cilia and flagella  as flexible active filaments~\cite{David,GueronMetachronal,GueronFirst,LironSim}.   
In order to help determine the exact physical process that, when  combined with hydrodynamics, leads  to synchrony, the   community has also made use of  minimal models of cilia where each individual  cilium is replaced by a small rigid sphere driven along a circular orbit~\cite{Rower,Lenz_2006,slovenci}. The simplicity of these sphere models not only allows for  analytical approaches but it  also makes the conclusions   robust to changes in the geometrical details.}
 
\begin{figure}[t]
   	\includegraphics[keepaspectratio,width=0.9\linewidth]{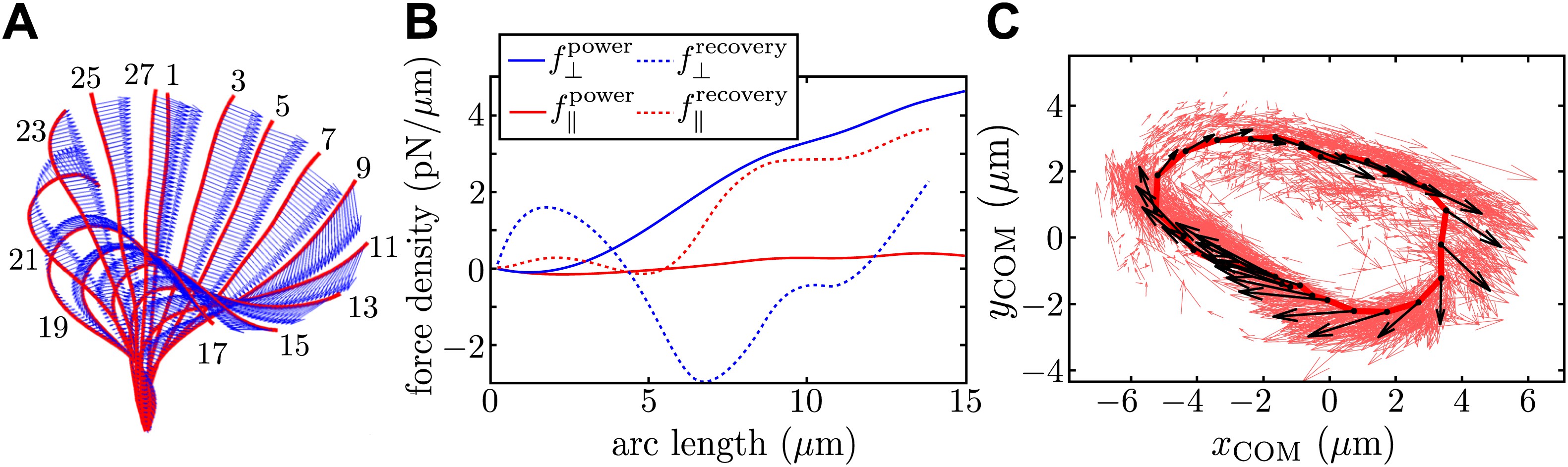}
   	\caption{Illustration of the dynamics of the ciliary beat pattern using an individual flagellum on the surface of the green alga {\it Volvox carteri}. A: Example of how the shape of a flagellum or cilium evolves during a beat. B:  Force density distribution along the flagellum or cilium during the power and recovery strokes. C: Location of the centre-of-drag of the flagellum or cilium  plotted together with the total force it exerts on the surrounding fluid. Reproduced with permission from Ref.~\cite{kirsty}, licensed under \href{https://creativecommons.org/licenses/by/4.0/}{CC BY 4.0}.}
	\label{fig:intro}
   \end{figure}

 Using these minimal models, two generic  physical mechanisms    have been identified as enabling hydrodynamic synchronisation, namely one relying on {\it elastic compliance} of the appendages and one requiring {\it  modulation of the force} driving the cilia. 
 Elastic compliance includes all features that originate from the bending elasticity  of the appendage (flagellum or cilium) or of its connection to the cell. It was demonstrated to be a relevant physical mechanism leading { to} hydrodynamic synchronisation in experiments~\cite{KennyExp}, numerical simulations~\cite{StokerFlexHelix} and theoretical calculations~\cite{niedermayer}.  In fact, elasticity  was long thought of as a requirement to induce hydrodynamic synchronisation until the force modulation mechanism was   proposed in Ref.~\cite{UchidaFM}. Developed specifically to address the dynamics of cilia, force modulation allows the forcing on the cilium to depend on the location of the cilium in its phase space, a model that  reflects the ability of an individual cilium to undergo an asymmetric two-phase beat~\cite{Blake_cilia}: a power stroke vigorously pushing the fluid followed by a recovery stroke hydrodynamically hiding with a smaller disturbance to  the fluid. Experiments with the flagella on the surface of the green alga {\it Volvox} and reproduced in  Fig.~\ref{fig:intro} allow one to illustrate this asymmetric time-varying deformation and the modulation of the force experienced by the cilium along its path.

The synchronisation of ciliary arrays is an important biological feature  and the lack  of synchronisation can lead to severe pathological conditions~\cite{diseases}.     One way to try and distinguish between  them, and potentially  uncover the physical mechanism{  most essential overall} in biology,  is to probe the robustness of these mechanisms under additional constraints.   
    Ciliary arrays are often found in confined areas of the body, such as narrow tubes or cavities inside more complex organisms. Examples include the  Fallopian tube of humans~\cite{fallopian} and the walls of the human brain ventricles~\cite{brain}. Confinement is also relevant for synchronisation  during the close encounter of ciliated organisms with each other or with flat surfaces, e.g.,~collisions of two swimming \textit{Paramecium} cells~\cite{paramecium_collisions}. In this paper, we therefore ask the following question: What  is the general impact of strong  geometric confinement  on the 
hydrodynamic synchronisation of cellular appendages? 
We use the minimal sphere models of interacting cilia  and, in order to model the effect of  strong hydrodynamic  confinement, place them between two nearby parallel rigid surfaces in the limit where  the distance between the surfaces is much smaller than the typical distance between the model cilia. We then investigate how each of the  two synchronisation mechanisms (elastic compliance vs.~force modulation) separately performs under confinement. We   compare our results  with the standard  case of model cilia above a single surface and show that, in the biologically-relevant situation of nodal cilia, the force modulation mechanism is significantly more robust to confinement than elastic compliance.{
We finally carry out a similar analysis on a second biologically motivated example, that  of primary cilia; in that case, the  difference  in robustness between the mechanism is not as pronounced but it appears to still favour  force modulation.}

Our paper is organised as follows. In Sec.~\ref{genset}, we outline the general mathematical setup and notation that is used throughout this study. In Sec.~\ref{midplane} we consider a preliminary (symmetric) example that demonstrates the differences between the mechanisms. The most general synchronisation dynamics, and its comparison with the standard one-wall case, is  investigated in Sec.~\ref{arbor} with 
 Sec.~\ref{EC} devoted to the performance of  elastic compliance and Sec.~\ref{FM} to force modulation. In Sec.~\ref{comparison} we apply our results to the{ cases of nodal and primary cilia} and address the robustness of the two mechanisms to confinement.  We finally summarise and discuss our results in  Sec.~\ref{dicussion}.

\section{Modelling cilia motion near surfaces\label{genset}}

\subsection{Minimal cilia model}

One of the first mathematical models for synchronisation of flagella was proposed by Taylor, who considered the interaction of a pair of   two-dimensional, infinite waving sheets~\cite{taylor}.{ In order to  examine analytically the collective behaviour of cilia under strong confinement, it is necessary to use instead a finite-size model whose dynamics can be subject to either one of the two physical mechanisms discussed above}. We thus adopt a version of the sphere model from Refs.~\cite{Lenz_2006,slovenci} in which individual cilia are modelled by small rigid spheres whose motion in the fluid represents the tip of a cilium, or its centre-of-drag,    driven along a cyclic orbit. 

This minimal model is supported by experimental measurements, some of which have been reproduced in Fig.~\ref{fig:intro} in the case of the green alga {\it Volvox}~\cite{kirsty}. The actual motion of an individual  cilium is recorded (technically it is termed a flagellum for this organism but its dynamics is very similar to that of standard cilia) (see Fig.~\ref{fig:intro}A). This allows one to estimate the instantaneous  velocity of the cilium  along its path and,  in conjunction with resistive-force theory for slender filaments \cite{GRAY_rft}, to predict the distribution of hydrodynamic forces along the cilium  (see Fig.~\ref{fig:intro}B). Using the predicted force densities, a centre-of-drag is found by weighting the arc-length positions with the corresponding force densities and the  evolution of the centre-of-drag is shown in Fig.~\ref{fig:intro}C together with the total drag force that the cilium exerts on the fluid. Examining the shape of its trajectory, it clearly supports the minimal model of a sphere driven along a cyclic orbit.

\subsection{Geometry}

 \begin{figure}[t]
	\includegraphics[width=\linewidth]{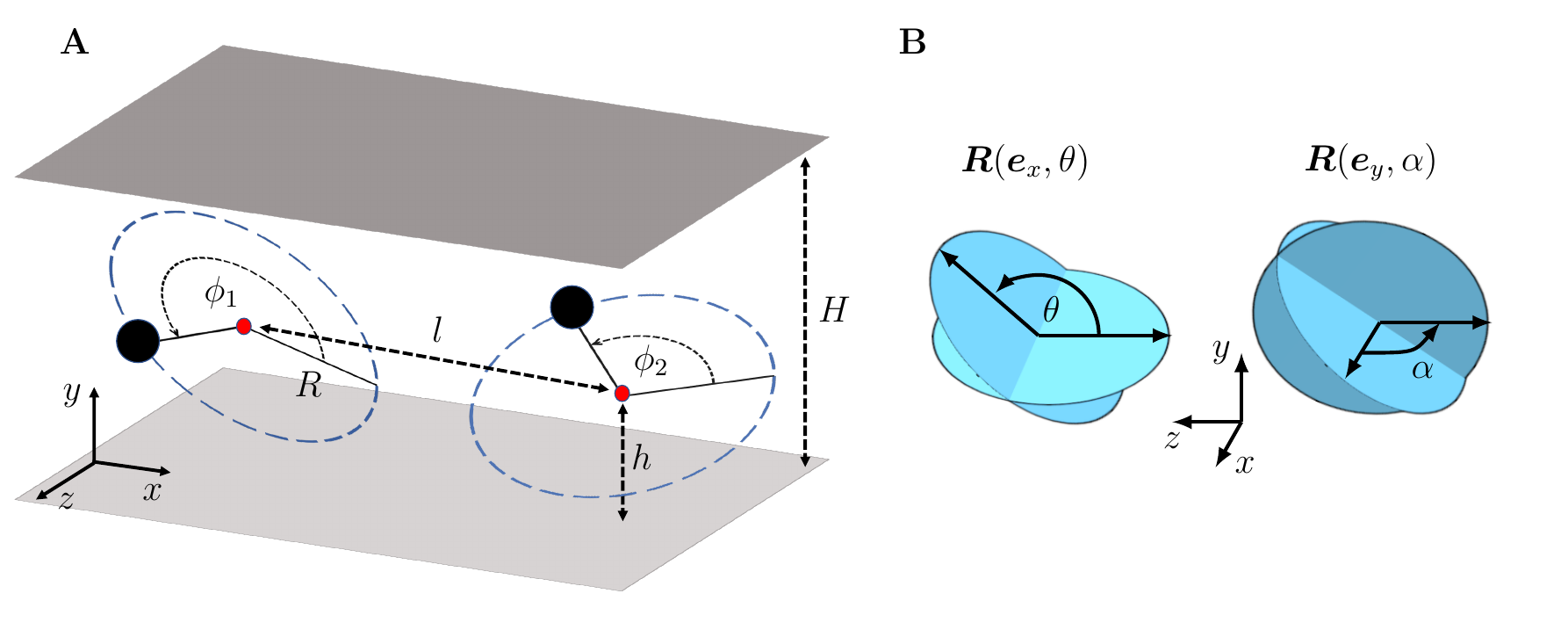}
		\caption{Illustration of the general geometrical setup used in our minimal model of cilia dynamics near walls.{ \bf A}: The relative position of the orbits and the walls. The distance between the walls $H$, the orbital radius $R$, the height of the centres above the lower wall $h$, the distance between the centres $l$ and the phase angles $\phi_1$ and $\phi_2$ are all clearly marked.{ \bf B}: Definitions of the angles $\theta$ and $\alpha$.}\label{setup}
\end{figure}

The general situation considered in this paper is therefore a pair of such spheres, each of radius $a$, forced along circular orbits  placed between two infinite walls that are parallel to each other, see Fig.~\ref{setup}. The entire gap between the walls is filled with an otherwise quiescent fluid of dynamic viscosity $\mu$. We will also consider the case where the upper wall is not present for the purpose of comparison~\cite{niedermayer}. Unless otherwise stated the top wall is at distance $H$ from the lower one.

The orbits, in general, will be allowed to have a variable radius $R_i, i=1,2$ but have a fixed centre at   height $h$ from the lower wall while their centres are set a distance $l$ apart (see notation in Fig.~\ref{setup}{ \bf A}). The orientation of a particular orbit is described by two angles, $\theta_i$ (termed the tilt) and $\alpha_i$ in such a way that the normal to the plane of motion is $\bm{n}_i =  \bm{R}(\bm{e}_y,\alpha_i)\cdot\bm{R}(\bm{e}_x, \theta_i)\cdot \bm{e}_y$ (see Fig.~\ref{setup}{ \bf B}){ where $\bm{R}(\bm{v},\varphi)$ is a matrix representing a rotation around the axis parallel to the vector $\bm{v}$ by the angle $\varphi$ in the positive mathematical sense and where $\bm{e}_v$ is a unit vector in the $\bm{v}$ direction}. We measure the phase $\phi_i$  such that the radius vector of the sphere is $\bm{e}_{R_i} =\bm{R}(\bm{e}_y,\alpha_i)\cdot\bm{R}(\bm{e}_x, \theta_i)\cdot\bm{R}(\bm{e}_y,\phi_i)\cdot\bm{e}_x = {\bf M}_i(\phi_i)\cdot \bm{e}_x$. The tangential direction, in which the sphere is forced, is described by a unit vector $\bm{e}_{\phi_i} = -{\bf M}_i(\phi_i)\cdot \bm{e}_z$. 

\subsection{Mechanics}

Following the calculations in Ref.~\cite{niedermayer} we model the elasticity of the cilium  as a linear restoring force that is imposed in the radial direction and that favours the unperturbed radius $R_0$. Spheres are forced along orbits by a tangential forcing $F(\phi)$, and the existence of power and recovery strokes is reflected in the explicit phase dependence of the forcing~\cite{UchidaFM}. 

Since our goal is to model motile appendages of microorganisms it is safe to assume that the fluid flow will be dominated by   viscous forces (typically $Re\lesssim 10^{-4}$~\cite{yates}) and thus described by the incompressible Stokes equations. In the Stokes limit, the spherical spheres must be force and torque free and we can use   Fax\'en's laws to account for the mutual interaction of the spheres to leading order~\cite{pozrikidis_1992}. 
The total drag on each sphere,  resulting from its motion in the flow created by the neighbouring sphere, is driven by the prescribed force in the tangential direction along the path and balanced by elasticity in the radial (perpendicular) direction~\cite{niedermayer,UchidaFM}. At leading order in $a^2/l^2$,   Fax\'en's first law for the sphere  marked with index 1 and projected onto the radial and tangential balance has the following form
\begin{eqnarray}
\zeta (R_1\dot{\phi_1}-\bm{e}_{\phi_1}\cdot \bm{u}_{2\rightarrow 1}) &=& F(\phi_1), \label{tangential} \\
\zeta (\dot{R}_1-\bm{e}_{R_1}\cdot \bm{u}_{2\rightarrow 1}) &=& -k(R_1-R_0), \label{radial}
\end{eqnarray}  
where $\zeta = 6\pi\mu a$ is the hydrodynamic drag coefficient of the sphere, $k$ is the spring constant of the radial elastic restoring force and $\bm{u}_{2\rightarrow 1}$ is the fluid velocity of the flow caused by the motion of sphere 2 evaluated at the centre of the sphere 1. 
Note that in a confined geometry, such as the one that we consider here, the hydrodynamic resistivity  of the sphere, $\zeta$, depends on the proximity of the sphere to either of the two walls. We will assume that the size of the sphere $a$ is much smaller than its distance to either of the walls and thus will neglect the corrections so that $\zeta$ takes  the value for a sphere in a bulk fluid. 

The radial equation, Eq.~\eqref{radial},  can then be simplified by further assuming  that the elasto-hydrodynamic relaxation time, $\tau = \zeta/k$, is much shorter than the time-scale on which hydrodynamic interaction changes the radius considerably. 
{ The latter time-scale increases when the interactions become weaker, i.e.,~when the spheres are further away from each other. Thus, we consider the far-field limit, $R\ll l$, and the quasi-steady approximation outlined above can then be justified \cite{niedermayer} by using experimental measurements of the physical parameters involved in the model~\cite{Camalet_2000,Blake_cilia}.}
  In practical terms, this means that the term  $\dot{R}_1$ will then be neglected in what follows, an assumption that can then be verified a posteriori.

\subsection{Hydrodynamic interactions}

In order to model the flow induced by each of the spheres, we   use the far-field approximation, thus assuming that the spheres are much further apart than their size, $a\ll l$. Having in mind the linearity of Stokes equations we can always write the flow field as $\bm{u}_{2\rightarrow 1} = {\bf G} (\bm{r}_1,\bm{r}_2)\cdot \bm{F}_2$, where $\bm{F}_2$ is the total external force applied to   sphere number 2 and ${\bf G}$ is a second rank tensor. Since the restoring elastic force is a response to mutual interaction of the spheres, to leading order in $a/l$ we have $\bm{F}_2\approx F(\phi_2) \bm{e}_{\phi_2}$. 
The flow caused by a moving sphere is forced by the surface stress distribution on the surface of the sphere through the boundary integral formula. In the  far-field, we then  asymptotically expand the surface integral in powers of the spatial decay of the flow. The leading-order term is   a point force singularity  resulting from  the total external force applied on the sphere. In an unbounded fluid, ${\bf G}$ is a stokeslet, which has the form of the Oseen tensor $(\mathbb{I}+\bm{r}\bm{r}/r^2)/8\pi\mu r$, with $\bm{r} = \bm{r}_1-\bm{r}_2$ where $\mathbb{I}$ is the identity tensor.

Under confinement, the solid surfaces will   screen hydrodynamic interactions. A standard way of dealing with hydrodynamic singularities near solid boundaries is to use   image systems, similar to the one in electrodynamics, where a set of hydrodynamic singularities is introduced in a fictional fluid domain within the solid boundary to balance the original  singularity within the fluid and enforce the no-slip boundary condition for the flow on the surface. Classically, it was shown by Blake \cite{blake_1971} that a stokeslet near a wall induces a set of three images, a \textquotedblleft
stokeslet equal in magnitude but opposite in sign to the initial stokeslet, a stokes doublet and a source-doublet, the displacement axes for the doublets being in the 
original direction of the force.\textquotedblright 

Adding a second wall significantly complicates the image system   and now an infinite series of hydrodynamic images with respect to each of the walls is required in order to enforce  the no-slip boundary condition   on both surfaces. Liron and Mochon \cite{Liron1976} managed to simplify the expression for the leading-order  flow field far from the stokeslet in the direction parallel to the surfaces, and obtained
\begin{equation}\label{eq:3}
\bm{u}_{1 \rightarrow 2} \approx \frac{3H}{2 \pi \mu l^2} \left(\frac{l}{\rho}\right)^2 \underbrace{\frac{y_1}{H}\left(1 -\frac{y_1}{H}\right)  \frac{y_2}{H}\left(1-\frac{y_2}{H}\right)}_{\displaystyle\langle y_1,y_2\rangle}\underbrace{(2\hat{\bm{\rho}}\hat{\bm{\rho}} - \mathbb{I}+\bm{e}_y \bm{e}_y)}_{\displaystyle{\bf S}}\cdot F(\phi_1) \bm{e}_{\phi_1},
\end{equation}
where $\bm{\rho} = (\mathbb{I}-\bm{e}_y \bm{e}_y)\cdot \bm{r}_{1\rightarrow2} = l\bm{e}_x +O(R\cos \theta /l)$ is the relative horizontal position of the two spheres and $y_i = h+R_i \sin \theta_i \sin \phi_i$ is the $y$ coordinate of the $i$th sphere. Note that we have defined the function $\langle y_1,y_2\rangle$ in Eq.~\eqref{eq:3} as well as the tensor $\bf S$. This far-field flow is exact up to  corrections that decay exponentially with the horizontal distance over the characteristic length-scale of the inter-wall distance, $H$. Thus, the Green's function for hydrodynamic interactions in this setting can be well approximated by 
\begin{equation}
{\bf G}(r_2,r_1) \approx \zeta^{-1} \beta \langle y_1,y_2\rangle l^2 \rho^{-2} \, {\bf S},
\end{equation}
 where $\beta \triangleq  9Ha/l^2$ is a dimensionless group that quantifies the strength of the flow due to sphere 1 near sphere 2 (or vice versa) compared with the unperturbed linear velocity of the spheres. 
 To simplify the notation, let us introduce the tensor ${\bf T} \triangleq \zeta \beta^{-1}\, {\bf M}_2^T\cdot {\bf G} \cdot {\bf M}_1$,   which is purely geometric and quantifies the interaction as
\begin{equation}
	\bm{e}_{R_2} \cdot \bm{u}_{1\rightarrow 2} \approx \beta \bm{e}_{x} \cdot {\bf M}_2^{T}\cdot{\bf G}\cdot(-{\bf M}_1) \cdot \bm{e}_{z}\, V_1 = -\beta T_{xz}\, V_1.
\end{equation}
and similarly
\begin{eqnarray}
	\bm{e}_{\phi_2} \cdot \bm{u}_{1\rightarrow 2} &\approx& \beta T_{zz}\, V_1, \nonumber \\
	\bm{e}_{R_1} \cdot \bm{u}_{2\rightarrow 1} &\approx& -\beta T^{T}_{xz} \, V_2 = - \beta T_{zx} \, V_2, \\
	\bm{e}_{\phi_1} \cdot \bm{u}_{2\rightarrow 1} &\approx& \beta T^{T}_{zz}\, V_2, \nonumber
\end{eqnarray}
where $V_i = F(\phi_i)/\zeta$ is the leading-order approximation of the instantaneous, phase dependent, tangential velocity of the $i$th sphere.

\subsection{Phase dynamics}\label{phase intro}

In the asymptotic limit where the  spheres are  far apart, i.e.~$\beta \ll 1$, and under the aforementioned assumption that the radial dynamics are quasi-steady, the radial Eq.~\eqref{radial} for the first sphere  yields $R_1 = R_0 - \beta \tau T_{zx} V_2$. Substituting $R_1$ into the phase equation, 
Eq.~\eqref{tangential}, we obtain 
\begin{equation} \label{step}
	\dot{\phi_1} = (V_1+\beta T_{zz} V_2)R_0^{-1}\left(1-\beta \tau T_{zx} \frac{V_2}{R_0}\right)^{-1}.
\end{equation}
After expanding the right hand side of Eq.~\eqref{step} to first order in $\beta$ and carrying out the same procedure for the second sphere, the phase evolution ends up being described by the  coupled system of nonlinear equations
\begin{eqnarray}\label{generaleq1}
\dot{\phi}_1 &=& \frac{F(\phi_1)}{\zeta R_0}+\beta\frac{F(\phi_2)}{\zeta R_0} T_{zz}  +\beta\tau\frac{F(\phi_1)}{\zeta R_0} \frac{F(\phi_2)}{\zeta R_0}T_{zx}, \\ \label{generaleq2}
\dot{\phi}_2 &=& \frac{F(\phi_2)}{\zeta R_0}+\beta\frac{F(\phi_1)}{\zeta R_0} T_{zz}  +\beta\tau\frac{F(\phi_2)}{\zeta R_0} \frac{F(\phi_1)}{\zeta R_0} T_{xz}.
\end{eqnarray}
For most of the calculations presented here it will be necessary to simplify the tensor ${\bf T}$ by truncating it at   leading order in $R/l$, and, under  this approximation, ${\bf S}$ simplifies   to $\bm{e}_x\bm{e}_x-\bm{e}_z\bm{e}_z$. 

In order to examine solely the effects of elastic compliance,  we set $F(\phi) = \zeta R_0\omega_0$ to be constant, where $\omega_0$ is the angular frequency of an isolated sphere. In contrast, in order to examine  solely the effects of force modulation, we set $k \to \infty$ (which implies $\tau = 0$).

\subsection{Synchronisation}
\label{sec:gauge}
Two oscillators are said to synchronise if, through mutual interaction, the phase difference $\phi_2-\phi_1$ converges to a value that is constant in time (a phase-locked state). However, when the forcing is phase dependent, the measured phase difference can be dominated by the difference in forcing and the final synchronisation states (other than completely in-phase) are less physically obvious from the information on  $\phi_2-\phi_1$. A way of solving this problem, without having to compare the contributions from the phase dependent forcing and the interactions, is to change below to a different phase gauge so that the phase dependent forcing effects are essentially factored out.{ Note that this new phase gauge is, in fact, how the phase is often defined in literature  for complex systems with periodic dynamics~\cite{pikovsky_rosenblum_kurths_2001}; due to the physical meaning of the  geometric phase $\phi$  introduced above, we choose to refer to it as a different phase gauge. Namely, it is denoted by $\Phi(\phi)$ and defined as} 
\begin{equation}\label{Phi}
\Phi(\phi) = 2 \pi \frac{K(\phi)}{K(2\pi)},\,\, \text{where} \,\,\,\, K(\phi) = \int_0^{\phi} \frac{d\phi^\prime}{F(\phi^\prime)},
\end{equation}
which evolves in time at a constant rate for an isolated sphere. Indeed, $\Phi(0) = 0, \Phi(2\pi) = 2\pi$, so $\Phi$ changes by $2\pi$ during a single revolution of the bead and, for an isolated sphere, we have
\begin{equation}
	\dot{\Phi} = \frac{2\pi}{K(2\pi)}\frac{\dot{\phi}}{F(\phi)} = \frac{2\pi}{K(2\pi)}(\zeta R_0)^{-1} \triangleq \Omega = const.
\end{equation}
Hence, this transformation is a proper change of phase gauge and,  for two spheres with the same orbit properties, the phase difference in the new gauge will depend only on the interactions. In situations where the force modulation model is examined, we will be interested in the time evolution of $\Delta \triangleq \Phi_1 - \Phi_2$. As we will usually be interested in the case where the force modulation is weak, i.e.,~the forcing varies only weakly from its mean value, the difference $\Phi(\phi) - \phi$ will be small, on the order of force modulation. Thus, if we reach a phase locked state $\Delta \equiv { \Delta_0}$ in the new gauge, we could say that $\phi_1-\phi_2 \sim { \Delta_0}$, asymptotically, with corrections on the order of the force modulation. As long as ${ \Delta_0}$ is of order $O(1)$ this is an asymptotically correct statement, which will allow us to conclude on the physics of synchronisation. However, $\Phi_1-\Phi_2 \equiv 0$ implies $\phi_1-\phi_2 \equiv 0$, so the asymptotic expression fails only for small but non-zero ${ \Delta_0}$, and this should be kept in mind when deriving physical conclusions.

\section{Cilia orbits in the mid-plane \label{midplane}}

In this section, we first consider the case where the spheres are orbiting in the plane located exactly half way between the two walls and parallel to both. This simple setup allows us to illustrate two important points: ($i$) the presence of the second (upper) wall has a markedly different impact on the elastic compliance and force modulation mechanisms, and ($ii$)  the relative orientation of the cilia is an important factor in both cases. Although the situation where the orbits are parallel to the walls is not a perfect representation of  ciliary beats, for any periodically-forced quasi-two-dimensional structure tightly fitting between plates, e.g.,~active artificial  microgels~\cite{microgel}, this is a  relevant limit to consider.
 
Setting the orbits in the mid-plane with $\alpha_i=\theta_i=0$ leads to a couple of mathematical simplifications. In that case, $\langle y_1,y_2 \rangle $ is exactly equal to $1/16$ so we will absorb it into $\beta =  9Ha/16l^2$, only in this section; furthermore,  ${\bf T}$ turns out to be a symmetric tensor even without approximating the interaction kernel ${\bf S}$. Knowing that $T_{xz} = T_{zx}$ makes it then obvious that a completely in-phase state $\phi_1=\phi_2$ is a solution of Eqs.~\eqref{generaleq1} and \eqref{generaleq2},   but to say the spheres synchronise in-phase we have to investigate the stability of this state.

\subsection{No synchronisation by elastic compliance}

First we address the case where the forcing along the orbit is constant, i.e.,~the elastic compliance model. Considering both Eqs.~\eqref{generaleq1} and \eqref{generaleq2}, with the terms $F(\phi_i)/\zeta R_0$ replaced by $\omega_0$, and since ${\bf T}$ is symmetric even without the linearisation of the interaction kernel in $R_0/l\ll 1$, the phase difference $\phi_1-\phi_2$ remains constant up to $O(\beta^2)$. This means that the elastic compliance mechanism leads to no synchronisation up to order $O(\beta^2)$. Comparing to the case of cilia orbits near a single wall \cite{niedermayer}, where the spheres always synchronise to an in-phase state at leading order in $\beta$, we thus see that in that the presence of strong hydrodynamic confinement disables elastic compliance as a synchronisation mechanism. 
 
 \subsection{Synchronisation by force modulation}

In the force modulation model (i.e.,~in the limit $\tau\rightarrow 0$), we can examine the linear stability of the in-phase state $\phi$ by introducing a small perturbation in the form $\phi_1 = \phi+\Delta, \phi_2 = \phi$, where 
\begin{equation}
	\dot{\phi} = \frac{F(\phi)}{\zeta R_0}\big[1+\beta T_{zz}(\phi,\phi)\big].
\end{equation}
 By calculating the average of the perturbation growth rate, $\langle \sigma \rangle$, and examining its sign over a single beat it is then possible to conclude on the stability characteristics of the in-phase state. The linearisation of Eqs.~\eqref{generaleq1} and \eqref{generaleq2} for small $\Delta$ leads to
\begin{equation}
	\dot{\Delta} = \frac{F(\phi+\Delta)-F(\phi)}{\zeta R_0}\big[1-\beta T_{zz}(\phi+\Delta, \phi)\big] = \frac{F^\prime(\phi)}{\zeta R_0}\big[1-\beta T_{zz}(\phi,\phi)\big]\Delta+O(\Delta^2),
\end{equation}
 which allows us to calculate the time-averaged growth rate as
 \begin{equation}
 	\langle \sigma \rangle \equiv \frac{1}{T_0} \int_{t}^{t+T_0} \frac{\dot{\Delta}}{\Delta}\,dt =  \frac{1}{T_0} \int_{0}^{2 \pi} \frac{F^\prime(\phi)}{F(\phi)}\frac{1-\beta T_{zz}}{1+\beta T_{zz}} \,d\phi = -\frac{2 \beta}{T_0} \int_{0}^{2 \pi} \frac{F^\prime(\phi)}{F(\phi)}T_{zz}\,d\phi+O(\beta^2),
 \end{equation}
 where $T_0$ is the period of the in-phase state $\phi$. These results turn out to be identical to those  obtained in the case of a single wall~\cite{UchidaFM} since both are characterised by $T_{zz}(\phi,\phi) =-C \cos 2\phi +const$, with $C>0$. A particular forcing function $F(\phi) $ that leads to in-phase synchronisation near a single wall will do so in the presence of the second one and vice versa, at order $\beta$. Thus, in this setting, the force modulation mechanism is   robust to strong hydrodynamic  confinement.

 \subsection{Impact of the relative orientation}

To illustrate the importance of the relative orientation of the orbits in this situation, we can consider the situation where one of them has been flipped by $180^\circ$, i.e.,~we let $\theta_2=\pi$. Physically this models a situation where the cilia are tethered on the opposing walls and thus they beat in different directions. In this case, the tensor ${\bf T}$ is no longer symmetric and the subtraction of Eqs.~\eqref{generaleq1} and \eqref{generaleq2} under   the elastic compliance model ($F(\phi)=\zeta R_0 \omega_0$) yields the equation for the phase difference $\Delta = \phi_1-\phi_2$ as
\begin{equation}
\dot{\Delta} \approx -2\beta \tau \omega_0^2 \sin \Delta,
\end{equation}
with corrections  of order $R/l$. This is the famous Adler equation with a stable equilibrium at $\Delta\equiv0$, and cilia  always synchronise to a completely in-phase state in that case. If the  case above of spheres beating in the same direction has indicated that elastic compliance might lead to no synchronisation in strong confinement, this shows that the relative orientation is an important factor that needs to be carefully included in any further investigation.

 \begin{figure}[t]
\includegraphics[width=0.85\textwidth]{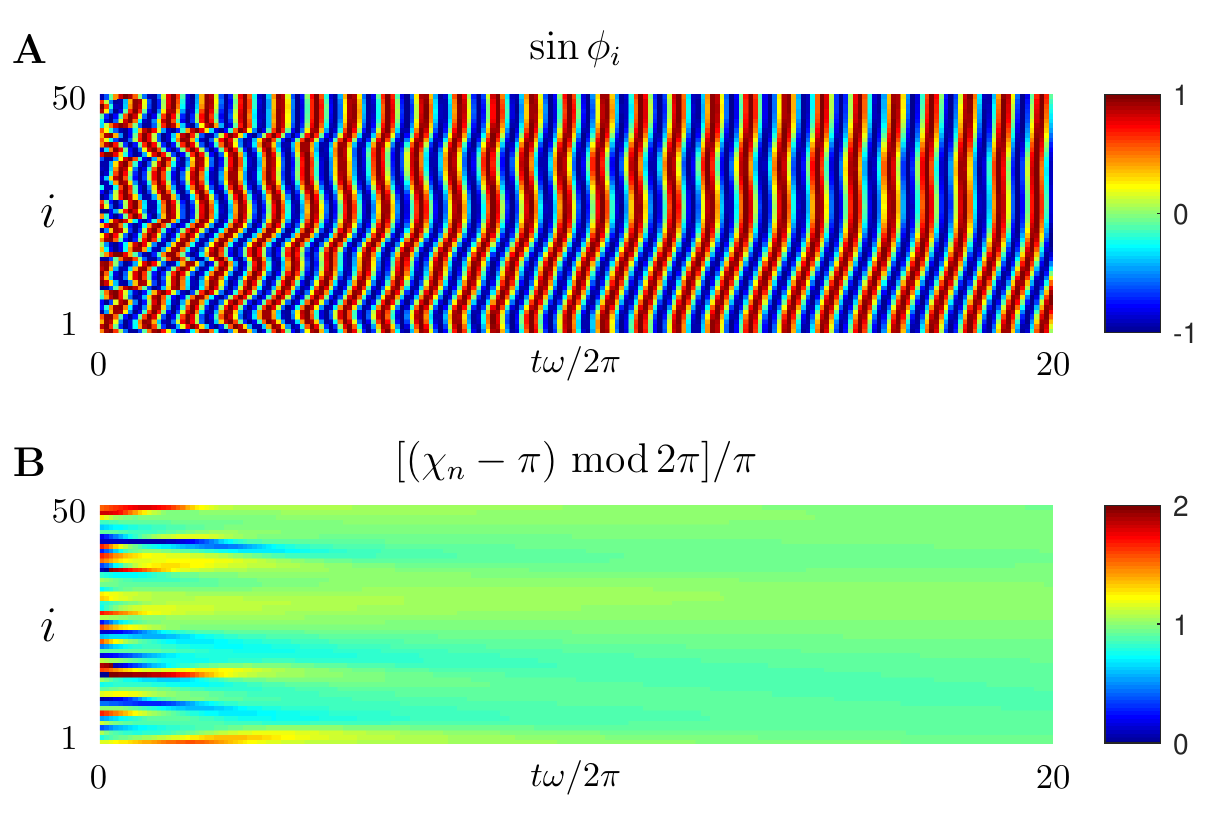}
	\caption{Formation of metachronal waves from random initial conditions with periodic boundary conditions for an array of cilia beating in alternating directions. We plot the time-evolution of $\sin \phi_i$ in{ subfigure {\bf A}} and $[(\chi_n-\pi)~{\rm mod}\,2\pi]/\pi$ in { subfigure {\bf B}} with $i=1,\ldots,N$. Simulations were done for $N = 50$ cilia with $\beta = 0.16$ and $\lambda = \tau \omega = 0.9$. Values for $\beta$ and $\lambda$ are chosen to be relatively large only for the purpose of reducing the number of periods the system takes for developing the metachronal waves.
	} \label{fig:metachronalsim}
\end{figure}

Beyond two-cilia synchronisation, elastic compliance can also lead to the development of metachronal waves for cilia beating in alternating directions. Consider a linear array of $2N$ model cilia, each at a distance $l$ from its neighbours and orbiting in the mid-plane such that $\theta_{2k} = 0$ and $\theta_{2k-1}=\pi$ for $k=1,\dots,N$.  We assume    that each sphere  interacts only with its   nearest neighbours on either side. Assuming periodic  boundary conditions where the spheres at the ends of the array   interact with each other, the  evolution equation  becomes
\begin{eqnarray}
	\dot{\phi}_n &=&\omega_0+\beta \omega_0 (\cos \chi_n+\cos \chi_{n+1})+\beta \tau \omega_0^2(\sin \chi_{n+1}-\sin \chi_n), \\
	\dot{\chi}_{n} &=& \beta \omega_0(\cos \chi_{n+1}-\cos \chi_{n-1})+\beta\tau \omega_0^2(\sin \chi_{n+1}-2 \sin \chi_{n} +\sin \chi_{n-1}), \label{metach}
\end{eqnarray}
where $\chi_{n} = \phi_n-\phi_{n-1}$. The same set of equations was reported in Ref.~\cite{niedermayer} for a chain of oscillators beating in the same direction near a single wall. Clearly, the system in Eq.~\eqref{metach} admits a solution $\chi_{n}=\Phi=const$, which corresponds to a metachronal wave where the neighbouring oscillators beat at the constant phase difference $\Phi$. This wave is unstable for $\cos \Phi <0$ and marginally stable for $\cos \Phi \geq 0$. We have verified the existence of these metachronal waves numerically for an array of 50 spheres with periodic boundary conditions and starting from random initial phases. Our results are illustrated numerically in Fig~\ref{fig:metachronalsim} in the form of the time evolution of the phases $\phi_n$ (Fig.~\ref{fig:metachronalsim}{\bf A}) and the phase differences of the neighbouring spheres $\chi_n$ (Fig.~\ref{fig:metachronalsim}{\bf B}). More precisely, $[(\chi_n-\pi)~{\rm mod}\,2\pi]/\pi$ is plotted as a proxy for $\chi_n$. Over time, the system reaches a metachronal state with $\chi_n$ close to $0$ for all $n=1,...,50$, which agrees with the prediction that the waves with $\cos \chi_n \geq 0$ are marginally stable. 

{This setup  is the simplest possible one to investigate  the organisation of many cilia and it shows that the mechanism is able to produce the same phenomenon (metachronal waves)   as in the single-wall case \cite{niedermayer}. Including full  long-range interactions beyond the nearest-neighbour case might further change the global organisation of cilia. It was previously shown that a similar minimal model exhibits near a single surface  a transition from metachronal waves to more complex states as the range of the interactions increases~\cite{chimera_transition}. The different nature of the Green's function in strong confinement, including the presence of the additional length-scale $H$, could of course lead to  different dynamics than in the single wall case.}

\section{Cilia with arbitrarily oriented orbits\label{arbor}}

As shown in the previous section, the orientation of the cilia orbits play a crucial role in allowing  synchronisation to take place. In this section, we therefore allow the orientations of the cilia arrays to be arbitrary and compute mathematically the consequence on the synchronisation dynamics for both physical mechanisms.  We will then apply our mathematical results to two biologically-relevant cases in Sec.~\ref{comparison}.

\subsection{Hydrodynamic synchronisation by elastic compliance \label{EC}}
In this first part,  we address the synchronisation by  elastic compliance. We use our general derivation in Eqs.~\eqref{generaleq1} and \eqref{generaleq2} and exclude the force modulation  by setting $F(\phi) = \zeta R \omega_0 = const$, which results in the   phase dynamics
\begin{eqnarray}
\dot{\phi}_1 &=& \omega_0 + \beta \tau \omega_0^2 T_{zx} + \beta \omega_0 T_{zz},\\
\dot{\phi}_2 &=& \omega_0 + \beta \tau \omega_0^2 T_{xz} + \beta \omega_0 T_{zz}.
\label{coupledinclinedgen}
\end{eqnarray}
By subtracting these two equations we obtain an equation for the phase difference $\Delta = \phi_1 - \phi_2$ as
\begin{equation}
\dot{\Delta} =\dot\phi_1-\dot\phi_2 = \beta \tau \omega^2 (T_{zx}-T_{xz}) .
\end{equation}
We can next  simplify ${\bf T}$ at   leading order in $R_0/l$ by approximating $l/\rho \approx 1$ and $\hat{\rho} \approx \bm{e}_x$, which leads to
\begin{eqnarray}\label{ECeqn}
\dot{\Delta}  \approx -\beta \tau \omega^2 \langle y_1, y_2 \rangle   \big[(\cos \theta_1 - \cos \theta_2) \sin (\alpha_1+\alpha_2) \cos \Delta +(1-\cos \theta_1 \cos \theta_2)\cos (\alpha_1+\alpha_2) \sin \Delta \big].  \qquad
\end{eqnarray}
{  The result in Eq.~\eqref{ECeqn} resembles the Adler equation,  a 
resemblance we can make more obvious using  a few algebraic manipulations.}  Defining $r_2\triangleq \big[(1-\cos \theta_1 \cos \theta_2)^2\cos (\alpha_1+\alpha_2)^2 + (\cos \theta_2 - \cos \theta_1)^2 \sin (\alpha_1+\alpha_2)^2\big]^{1/2}\geq0$ and the angle  $\Delta_0^{(2)}$ as 
\begin{eqnarray}
r_2 \cos \Delta_0^{(2)} &=& (1-\cos \theta_1 \cos \theta_2)\cos (\alpha_1+\alpha_2), \\
r_2 \sin \Delta_0^{(2)} &=&  (\cos \theta_2 - \cos \theta_1) \sin (\alpha_1+\alpha_2), \label{sin2}
\end{eqnarray} 
 transforms Eq.~\eqref{ECeqn} into 
\begin{equation}\label{EC_Adler}
	\dot{\Delta} = -\beta \tau \omega^2 \langle y_1,y_2 \rangle r_2 \sin (\Delta-\Delta_0^{(2)}).
\end{equation}
Note that we have used the superscript $2$ in the notation $\Delta_0^{(2)}$ in order to emphasise that this is the result in the case of two walls. 
Since all the prefactors in Eq.~\eqref{EC_Adler}, including $r_2$ and $\langle y_1,y_2\rangle$, are positive numbers, the phase difference $\Delta_0^{(2)}$ is the linearly stable solution to Eq.~\eqref{EC_Adler} while $\Delta_0^{(2)}+\pi$ is the  unstable solution. This   is true even though the term $\langle y_1,y_2 \rangle$ is, in fact, phase dependent and therefore   
Eq.~\eqref{EC_Adler} is not entirely equivalent to the Adler's equation.  It is important to note that both $r_2$ and $\Delta_0^{(2)}$ are a function of the orientation of the orbits only.{ Note that the     prefactor $r_2$ indicates that the magnitude of the contribution of the hydrodynamic interactions  to the rate of change of the phase difference strongly depends on the orientation of the orbits. It  is therefore an important factor in quantifying the strength of synchronisation towards $\Delta_0^{(2)}$ and thus in measuring the importance of hydrodynamic interactions  relative to any other  effects not included in our model, for example noise of any kind. } 
 
We can now compare these results with the case of a single wall. Past work involving solely elastic compliance as the synchronisation mechanism has  considered only the horizontal orbits,    so for the purpose of comparison we adjust the model of Ref.~\cite{niedermayer} to  arbitrarily oriented orbits close to a single wall. We obtain that result by changing the interaction kernel to a linearised Blake tensor $\tilde{\beta}\bm{e}_x\bm{e}_x$. Here, $\tilde{\beta} \triangleq 9ah^2/l^3$ and it is a parameter analogous to $\beta$ but relevant to the single wall case. These non-dimensional groups are  a relative measure of the speed of synchronisation since the number of periods around the orbits required before reaching a synchronised state is proportional to $1/\beta$. 

After performing the same calculations as for the case of two walls, we obtain
\begin{eqnarray}
\dot{\Delta} = \tilde{\beta} \tau \omega^2 (T_{zx}-T_{xz})
\approx -\tilde{\beta} \tau \omega^2 \bigg[(\cos \theta_1 \sin \alpha_1 \cos \alpha_2-\cos \theta_2 \sin \alpha_2 \cos \alpha_1) \cos \Delta \nonumber \\* 
+(\cos \alpha_1 \cos \alpha_2 + \sin \alpha_1 \sin \alpha_2 \cos \theta_1 \cos \theta_2) \sin \Delta \bigg],
\end{eqnarray}
and we see synchronisation similar to the case of two walls. Defining $r_1 > 0$ as
\begin{eqnarray}
r_1^2 &=& (\cos \theta_1 \sin \alpha_1 \cos \alpha_2-\cos \theta_2 \sin \alpha_2 \cos \alpha_1)^2 \nonumber \\ &+& (\cos \alpha_1 \cos \alpha_2 + \sin \alpha_1 \sin \alpha_2 \cos \theta_1 \cos \theta_2)^2, 
\end{eqnarray}
and the angle $\Delta_0^{(1)}$ via
\begin{eqnarray}
r_1 \cos \Delta_0^{(1)} &=& \cos \alpha_1 \cos \alpha_2 + \sin \alpha_1 \sin \alpha_2 \cos \theta_1 \cos \theta_2, \\*
r_1 \sin \Delta_0^{(1)} &=&  -\cos \theta_1 \sin \alpha_1 \cos \alpha_2+\cos \theta_2 \sin \alpha_2 \cos \alpha_1, \label{sin1}
\end{eqnarray} 
we obtain 
\begin{equation}
\dot{\Delta} = -\tilde{\beta} \tau \omega^2 r_1 \sin (\Delta-\Delta_0^{(1)}).
\end{equation}
Notice also that there is no synchronisation to leading order if and only if $r_1=0$. 
 It is notable that this condition{ can arise} in the case of identical orbits $\alpha_i = \alpha_0$ and $\theta_i = \theta_0$. 
 In this case, $r_1=0$  
  is equivalent to $\alpha_0 = \pm \pi/2$ and $\theta_0 = \pm \pi/2$, which is the case of vertical orbits perpendicular to the $x$ axis, i.e.,~to the direction connecting its centres. Intuitively this is clear since we linearised the Blake tensor to $\tilde{\beta} \bm{e}_x \bm{e}_x$, so in this case orbits have no interaction to this order.
  
  Thus, we have found that in both cases, one or two walls, the system synchronises to a phase locked state. The value of that terminal phase difference $\Delta^{(i)}_0$ (where the label $i$ is the number of walls in the setup) is different in cases of one and two walls and it is also a function of the orientation of the orbits. In  Sec.~\ref{comparison}, we will quantify the difference between $\Delta^{(1)}_0$ and $\Delta^{(2)}_0$ in two biologically-relevant cases and use this information as a measure of robustness to confinement of the elastic compliance as a mechanism for hydrodynamic synchronisation.
 
\subsection{Hydrodynamic synchronisation by force modulation \label{FM}}

In order to address the force modulation model, we follow the approach outlined   in 
Sec.~\ref{sec:gauge} and  perform a change of variable, $\Phi(\phi)$, so that the new phase variable now grows in time at constant rate $\Omega = 2 \pi/T_0$, where $T_0$ is the oscillation  period of an  isolated model cilium. The change of variables is given by
\begin{equation}\label{phi}
\Phi(\phi) = 2 \pi \frac{K(\phi)}{K(2\pi)}, \quad\text{where} \quad K(\phi) = \int_0^{\phi} \frac{d\phi^\prime}{F(\phi^\prime)},
\end{equation}
or, equivalently, $\dot{\Phi} = \dot{\phi}/F(\phi)$.
If we now let $\Phi_i = \Phi(\phi_i)$ and introduce $\bar{\Phi} = \frac{1}{2}(\Phi_1+\Phi_2)$ and $\Delta = \Phi_1-\Phi_2$, we can anticipate  that  $\Delta$ is a slow variable and $\Delta = 0$ is an in-phase solution. In general, it is possible to invert Eq.~\eqref{phi} so that $\phi_i = \phi(\Phi_i) = \phi(\bar{\Phi}\pm\Delta/2)$ and the phase evolution Eqs.~\eqref{generaleq1} and \eqref{generaleq2} become
\begin{eqnarray}
	 \frac{\dot{\Phi}_1}{\Omega} &=& 1 + \beta \frac{F(\Phi_2)}{F(\Phi_1)} T_{zz}(\Phi_1,\Phi_2), \\
	 \frac{\dot{\Phi}_2}{\Omega}&=& 1 + \beta \frac{F(\Phi_1)}{F(\Phi_2)} T_{zz}(\Phi_1,\Phi_2), 
\end{eqnarray}
where we have used the  shorthand  notation $f(\Phi) \equiv f(\phi(\Phi))$.
 Subtracting these two equations leads to 
 \begin{equation}\label{genDelta}
 	\dot{\Delta} = \beta\Omega \left[ \frac{F(\bar{\Phi}-\Delta/2)}{F(\bar{\Phi}+\Delta/2)}- \frac{F(\bar{\Phi}+\Delta/2)}{F(\bar{\Phi}-\Delta/2)}\right] T_{zz}(\bar{\Phi}+\Delta/2,\bar{\Phi}-\Delta/2) \equiv W(\bar{\Phi},\Delta).
 \end{equation}
  Under the assumption of weak interaction $\beta \ll 1$, we confirm a posteriori that  $\Delta$  is indeed a slow variable and $\bar{\Phi}\approx \Omega t$ at leading order in $\beta$.

  In order  to examine the evolution of the phase difference, it is instructive to coarse-grain the timescale by averaging Eq.~\eqref{genDelta} over the fast period of the mean phase, as
	 \begin{equation}
	\langle \dot{\Delta} \rangle = \frac{1}{2\pi}\int_0^{2\pi} W(\bar{\Phi},\Delta)\, d\bar{\Phi} \equiv -\frac{dV(\Delta)}{dt},
	\end{equation}   
where we have defined the effective potential that governs the long-time evolution of $\Delta$.{ The location of the global minimum of $V$, if it existed,  would be the state to which the system synchronises;  the goal is therefore to determine this potential. However, since $W(\bar{\Phi},\Delta)$ is a periodic function in $\Delta$, the function $V(\Delta)$ must have the form of a tilted washboard potential with no global minimum. A real physical system would thus spend a relatively long time in the local minima with occasional biased phase slips to the other minima caused by the stochastic noise (not included in our model). This feature has been observed in experiments~\cite{kirsty} and is correctly captured by the force modulation model~\cite{UchidaFM}. Hence, the relevant features of the potential $V(\Delta)$ for the synchronisation are (i) the location of local minima inside the  interval $[-\pi,\pi[$ and (ii) the direction of the tilt.

Since it is difficult to proceed further with a general force profile,} we will focus the first mode of force modulation, $F(\phi) = F_0[1+A \sin (\phi+\delta)]$. This choice is   motivated by experimental results on the dynamics of short flagella on the somatic cells of {\it Volvox carteri}~\cite{kirsty}. As shown in Fig.~\ref{fig:intro},   a hydrodynamic point force was fitted to  measured flow field from beating cilia; the magnitude of this fitted force shows one strong peak during a single beat, suggesting that the fundamental mode of force modulation is dominant.

The assumption that $A\lesssim 1$ allows us to invert Eq.~\eqref{phi} up to $O(A^2)$ and obtain
\begin{eqnarray} \label{inverse}
	\phi(\Phi) &=& \Phi +\big[\cos \delta - \cos (\Phi+\delta)\big]A \\  &+&\left\{\sin(\Phi+\delta)\big[\cos \delta - \cos(\Phi+\delta)\big]  +\frac{1}{4}\big[\sin (2\Phi+2\delta)-\sin 2\delta\big]\right\}A^2 +O(A^3).   \nonumber
\end{eqnarray}
Furthermore, in the process of recovering the effective potential $V(\Delta)$ the terms are truncated at $O(A^3)$ since, based on the single wall case \cite{UchidaFM}, we expect synchronisation at that order. We also approximate the hydrodynamic interactions at leading order in $R/l$ so that
\begin{eqnarray} \label{Tzz}
	T_{zz}(\phi_1,\phi_2) &\approx&  \langle y_1,y_2 \rangle \, \big\{\cos(\alpha_1+\alpha_2) \big[\sin \phi_1 \sin \phi_2 -\cos \theta_1 \cos \theta_2 \cos \phi_1 \cos \phi_2\big] \\ \nonumber
	&+& \sin(\alpha_1+ \alpha_2) \big[\cos \theta_1 \cos \phi_1 \sin \phi_2+\cos \theta_2 \cos \phi_2 \sin \phi_1\big]\big\}.
\end{eqnarray}

\begin{figure}[h]
	\includegraphics[width=0.9\textwidth]{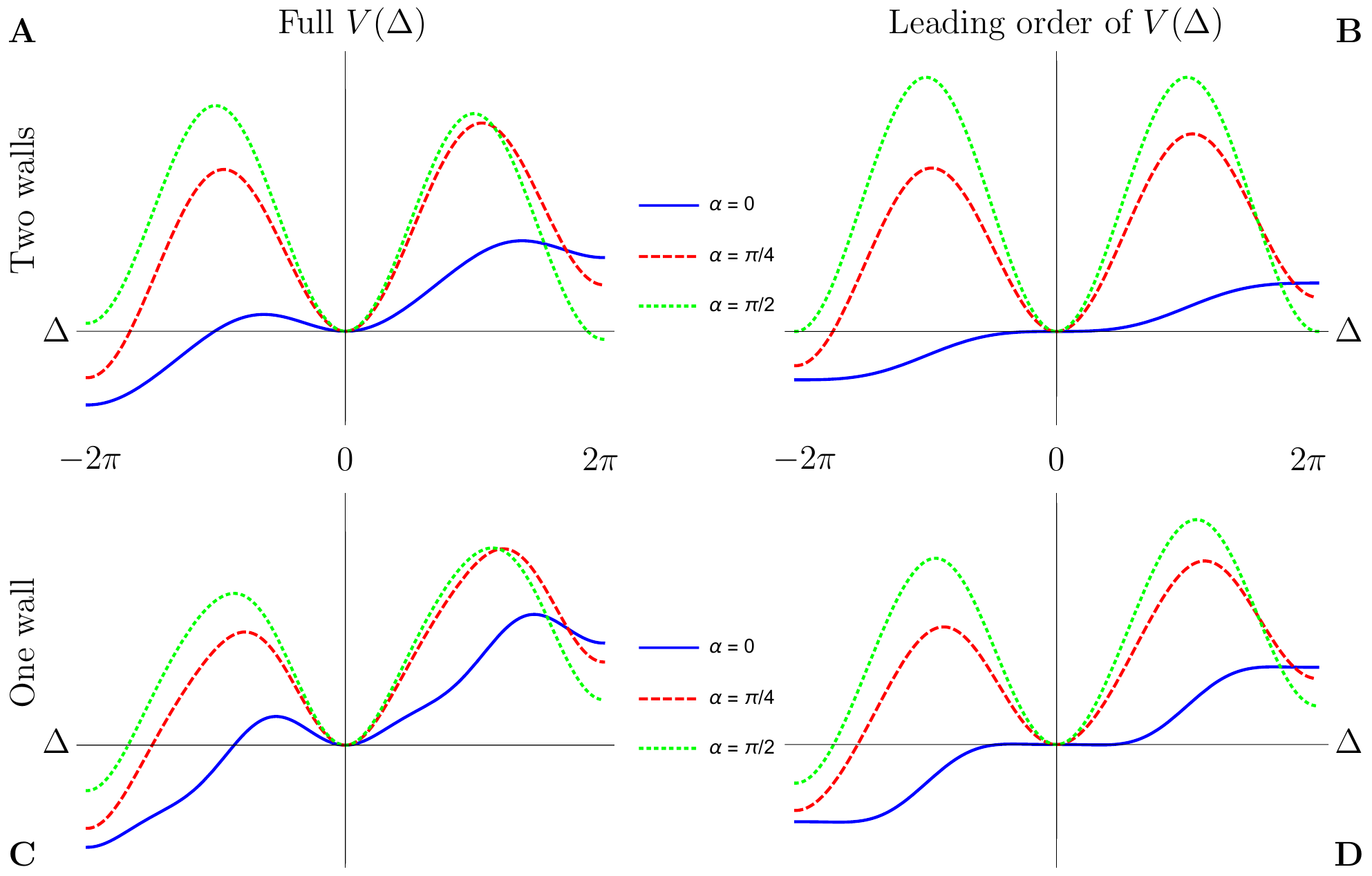}
	\caption{Effective synchronisation potential $V(\Delta)$ for three different values of $\alpha = (\alpha_1+\alpha_2)/2 = 0, \pi/4, \pi/2$. 
	Top row: Two-wall case, full potential  $V(\Delta)$ (in~{\bf A}) and its leading-order approximation in $h/H$ and $R/H$ (in~{\bf B}) for $\theta_1=0.4$, $ \theta_2=0.45$, $A=0.1$, $R = h = 0.1 H$, and $\delta=0$. Bottom row: Analogous potentials for the single-wall case ({\bf C, D} corresponding to~{\bf A, B}, respectively). Note the vertical axes are scaled differently, which does not affect the  location of the local minima in each graph. 
	}\label{potentials}
\end{figure}

The derivation of Eq.~\eqref{Tzz} is the last step  in this section that can be carried out by hand, while in what follows we employ symbolic manipulation  using Mathematica. Due to a large number of parameters of this model, even with the simplification of working to second order in $A$, the potential $V(\Delta)$ ends up having 5131 term and therefore we omit its full form here. We can proceed by making the further assumption that $h$ and $R$ are of  comparable magnitude. This  is supported by by experimental measurements of the cilium beating pattern \cite{Buceta2005} and by the fact that a sphere in the minimal cilium model    is meant to represent the tip of a cilium \cite{Lenz_2006}, so both the size of the orbit and the height above the surface  are expected to have the same scaling. Keeping only the leading-order terms in $h/H$ and $R/H$, we then obtain 
\begin{eqnarray}
	V(\Delta) &=& \beta \Omega A \frac{hR}{H^2} \left( C_0 \Delta + B_1 (1-\cos \Delta) + \frac{B_2}{2} (1-\cos 2\Delta) + A_1 \sin \Delta + \frac{A_2}{2} \sin 2\Delta \right) \nonumber \\ 
	&& \hspace{7cm}+ O(A^2 (h/H)^2,A (h/H)^3), \label{Potential2}
\end{eqnarray}
where the five constants are now defined as
\begin{eqnarray}
	C_0 &=&  -\frac{1}{2} \bigg[\sin 2\alpha \sin \delta \sin (\theta_1-\theta_2)+\cos 2\alpha \cos \delta
	(\sin \theta_1-\sin \theta_2)\bigg], \\
	B_1 &=& -\frac{1}{8} \bigg\{2 \cos 2\alpha \sin \delta (\sin {\theta_1}+\sin {\theta_2}) (\cos
	\theta_1 \cos \theta_2+1) \nonumber \\ 
	&& \hspace{4.2cm}+\sin 2\alpha \cos \delta [-6 \sin
	({\theta_1}+{\theta_2})+\sin 2 {\theta_1}+\sin 2 {\theta_2}]\bigg\}, \\
	B_2 &=& -\frac{1}{4} \bigg\{\cos 2\alpha \sin \delta (\sin \theta_1+\sin \theta_2) (1-\cos
	\theta_1 \cos \theta_2) \nonumber \\ 
	&& \hspace{4.2cm}+\sin 2\alpha \cos \delta [1-\cos(\theta_1-\theta_2)] \sin
	({\theta_1}+{\theta_2})\bigg\}, \\
	A_1 &=& -\frac{1}{4} (\sin \theta_1-\sin \theta_2) [\cos 2\alpha \cos \delta (\cos
	\theta_1 \cos \theta_2-3)-\sin 2\alpha \sin \delta (\cos \theta_1+\cos
	\theta_2)], \\
	A_2 &=& \frac{1}{4} \bigg\{\sin 2\alpha \sin \delta \sin ({\theta_1}-{\theta_2})[1-\cos (\theta_1+\theta_2)] \nonumber \\ 
	&& \hspace{4.2cm} +\cos 2\alpha \cos \delta (\sin
	\theta_1-\sin \theta_2) (\cos \theta_1 \cos \theta_2-1)\bigg\},
\end{eqnarray}
and $2\alpha = \alpha_1+\alpha_2$. In Fig.~\ref{potentials}, we plot  the full potential $V(\Delta)$ and its leading-order approximation, Eq.~\eqref{Potential2},  for $A=0.1$ and $R=h= 0.1H$ (top row). 
Note that since our focus is on the landscape, and specifically  the location of the local minima, we do not need to include the prefactors when plotting the leading order in  Fig.~\ref{potentials},  right.  
Even for the relatively large values of the parameters  treated as small in the calculations, we see that the landscape of the leading-order approximation agrees well with the full expression.

For the purposes of comparison, we can then repeat the calculations in the case of a single wall;  to best of our knowledge, the effective potential arising  for arbitrarily oriented orbits  has not been reported in the literature, though  the potentials were previously calculated and plotted for a few cases with $\theta_1=\theta_2$ in \cite{Uchida2012}. The only adjustment necessary in the calculations consists in altering the interaction kernel, where we now use the  Blake tensor linearised in $R/l$, $\zeta^{-1} \tilde{\beta}h^{-2} y_1 y_2\bm{e}_x \bm{e}_x$. Assuming again that the radii of the orbits and their distances to the substrate are comparable, we obtain in the one-wall case 
\begin{equation}
T_{zz} = \frac{y_1 y_2}{h^2} \big(\cos \phi_1 \cos \theta_1 \sin \alpha_1 +\cos \alpha_1 \sin \phi_1 \big) \big(\cos \phi_2 \cos \theta_2 \sin \alpha_2+\cos \alpha_2 \sin \phi_2 \big),
\end{equation}
with $\tilde{\beta}= 9ah^2/l^3$, which is   the same as previously defined in elastic compliance subsection. Finally, the potential ends up having the exact same form as Eq.~\eqref{Potential2} in the case of two walls but with a different prefactor of $\tilde{\beta} \Omega A R/h$ and with the constants now defined as
\begin{eqnarray}
C_0 &=&  \frac{1}{2} \big[\cos \alpha_2 \sin \theta_2 (\sin \alpha_1 \sin \delta \cos
\theta_1+\cos \alpha_1 \cos \delta) \nonumber \\
&& \hspace{4.2cm}-\cos \alpha_1 \sin \theta_1
(\sin \alpha_2 \sin \delta \cos \theta_2+\cos \alpha_2 \cos \delta)\big], \\
B_1 &=& \frac{1}{4} \bigg\{\sin \alpha_2 \cos \theta_2 [\cos \alpha_1 \cos \delta (3 \sin
\theta_1-\sin \theta_2)+\sin \alpha_1 \sin \delta \cos \theta_1
(\sin \theta_1+\sin \theta_2)]\nonumber \\ 
&&-\cos \alpha_2 [\sin \alpha_1 \cos
\delta \cos \theta_1 (\sin \theta_1-3 \sin \theta_2)+\cos \alpha_1
\sin \delta (\sin \theta_1+\sin \theta_2)]\bigg\}, \\
B_2 &=& \frac{1}{4} \bigg\{\cos \alpha_2 [\sin \alpha_1 \cos \delta \cos \theta_1 (\sin
\theta_1-\sin \theta_2)-\cos \alpha_1 \sin \delta (\sin
\theta_1+\sin \theta_2)] \nonumber \\ 
&& -\sin \alpha_2 \cos \theta_2 [\cos
\alpha_1 \cos \delta (\sin \theta_1-\sin \theta_2)+\sin \alpha_1
\sin \delta \cos \theta_1 (\sin \theta_1+\sin \theta_2)]\bigg\}, \\
A_1 &=& \frac{1}{4} (\sin \theta_1-\sin \theta_2) \big[\sin \alpha_1 \cos
\theta_1 (\sin \alpha_2 \cos \delta \cos \theta_2+\cos \alpha_2 \sin
\delta)\nonumber \\
&& \hspace{4.2cm} +\cos \alpha_1 (\sin \alpha_2 \sin \delta \cos \theta_2+3 \cos
\alpha_2 \cos \delta)\big], \\
A_2 &=& \frac{1}{4} \bigg\{\cos \alpha_1 [\cos \alpha_2 \cos \delta (\sin \theta_2-\sin
\theta_1)+\sin \alpha_2 \sin \delta \cos \theta_2 (\sin
\theta_1+\sin \theta_2)] \nonumber \\
&& -\sin \alpha_1 \cos \theta_1 [\sin
\alpha_2 \cos \delta \cos \theta_2 (\sin \theta_1-\sin
\theta_2)+\cos \alpha_2 \sin \delta (\sin \theta_1+\sin \theta_2)]\bigg\}.
\end{eqnarray}
Similarly to the case of two walls, this leading-order approximation in $h/H$ an$R/H$ agrees well with the full form for relatively large force modulation amplitude, as   illustrated in  Fig.~\ref{potentials}. Compared with previous work \cite{Uchida2012,Maestro2018}, it is notable that the leading-order   potential is now linear in $A$ since in these references it was reported to be quadratic. This is because past work only considered the case where the orbits were horizontal, $\theta_1=\theta_2=0$;   in that case, the leading order reported here  vanishes, thereby leading to synchronisation at order $O(A^2)$.

A limit that is easier to compare with  previous work is when  the vertical amplitude of motion, $R \sin \theta_i$, is assumed to be much smaller  than the height $h$ of the centre of the orbit from the bottom wall. This assumption, that does not appear  very relevant in the context of biology, is, however, what is usually done in the previous toy models~\cite{niedermayer,Uchida2012}. Assuming $R \sin \theta_i \ll h$, we can then approximate $\langle y_1,y_2\rangle \approx h^2/H^2$ and after repeating the aforementioned procedure in the case of a single wall and in the case of the two walls, we obtain
\begin{equation}
\begin{aligned}
\frac{V_1(\Delta)}{\tilde{\beta}\Omega} = \frac{A^2}{4}(1-\cos \Delta)r_1(\pi-\theta_1,\theta_2) \sin \big[2\delta-\Delta_0^{(1)}(\pi-\theta_1,\theta_2)\big], \\
\frac{V_2(\Delta)}{\beta\Omega} = \frac{A^2}{4}(1-\cos \Delta)r_2(\pi-\theta_1,\theta_2) \sin \big[2\delta-\Delta_0^{(2)}(\pi-\theta_1,\theta_2)\big].
\end{aligned}\label{FMpotentials}
\end{equation}
where $V_i$ is relevant for the case with $i$ walls and $r_i$ and $\Delta^{(i)}$ are the same as in the elastic compliance model but with the substitution $\theta_1 \mapsto \pi-\theta_1$. Again, $\tilde{\beta}$ is the same as previously defined above.

Just like the elastic compliance mechanism,  force modulation therefore always leads   to a single stable phase-locked state  both for one and two walls. To examine the robustness of force modulation to strong hydrodynamic confinement, we examine in the next section   the differences  between the phase-locked states in two biologically-relevant cases, and compare them to  elastic compliance.

{
\section{Robustness of synchronisation mechanisms \label{comparison}}
}
Since there are many degrees of freedom even in the minimal models we consider, it is  difficult to compare the results obtained for different synchronisation mechanisms across the entire parameter range. Furthermore, it is clear that some of the parameter values would be    far from any of the biologically-relevant limits that  motivated this work originally. In this section, we therefore focus on{ two} specific  biological situations in order to set biologically-relevant values for the parameters in our model.

{ Before  introducing the details of the  biological examples,  we can already fix some of the model parameters by using the general knowledge of  of ciliary arrays.} First, cilia are known to self organise to all beat in the same direction~\cite{Joanny, exp_reorient}. Although hydrodynamic interactions could in theory play a role in setting this, it is a reasonable assumption that we will assume to be  true in what follows. Using our notation, this means that $\alpha_1=\alpha_2 = \alpha$. Secondly, the physical interpretation of the phase shift $\delta$ in the assumed force profile that we have used in our calculations, $F(\phi) = F_0[1+A \sin(\phi+\delta)]$, is that the maximal drag on the cilium will occur when $\phi = \pi/2-\delta$. Although it is rather difficult to measure this phase of maximal force, in the model that we are using here, $\phi = \pi/2$ represents the point at which the sphere is at its highest distance from the surface on which the  cilium is attached. This means that at this phase, the sphere is most exposed to the fluid drag and it is thus reasonable to assume that  $\delta = 0$.

{ 
 As   demonstrated above, both physical mechanisms (elastic compliance vs.~force modulation) lead to stable phase-locked states. 
In order to quantify the  difference in the synchronised states between the  presence of two walls (strong confinement) and that of a single nearby surface, we then introduce the  normalised distance   $\mu_{1-2}$ between  the 	``locked-in'' phases on a unit circle,
$\mu_{1-2} = (|\Delta_0^{(2)}-\Delta_0^{(1)}|\mod \pi)/\pi$. Thus, $\mu_{1-2}=0$ means that under both levels of confinement these locked-in phases are identical  while if they differ by the maximal value of $\pi$ the distance will take its maximal value of $\mu_{1-2}=1$. For the elastic compliance mechanism, the value of $\Delta_0^{(i)}$, with  $i=1,2$ was previously defined as a stable fixed point of the phase difference. Similarly, in the context of force modulation, this definition extends to a stable fixed point obtained as a local minimum of $V(\Delta)$ inside the interval $[-\pi,\pi[$.

We now consider separately two common types of cilia, nodal and primary, due to their different geometrical features.
}

 \subsection{Nodal cilia}

\begin{figure}\centering
 \includegraphics[width=0.7\linewidth]{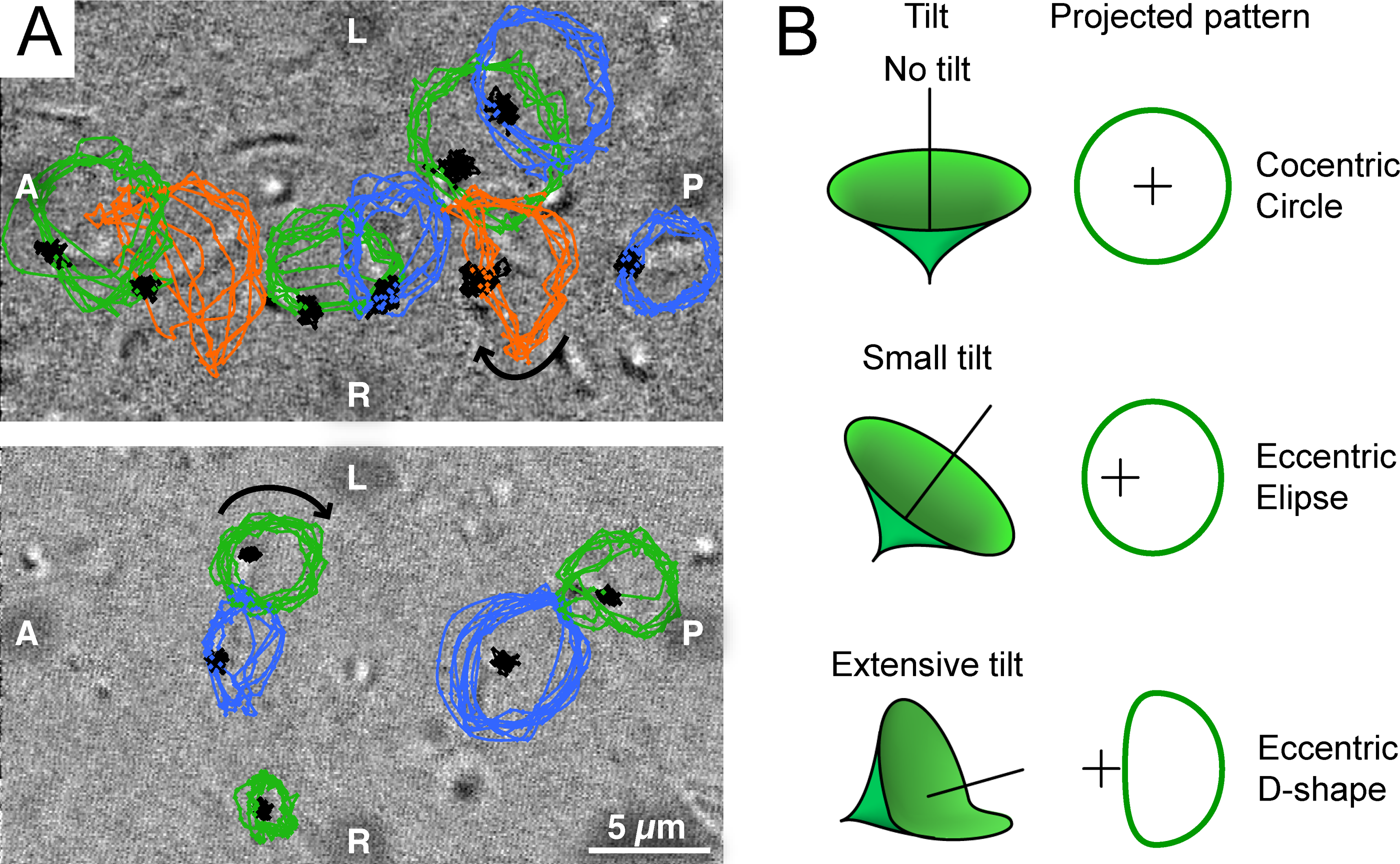}
	\caption
{
Dynamics of nodal cilia in mice embryos. Experimental measurements of the   trace of the cilium  tip motion (left)  are used to determine the  average tilt of nodal cilia (right). Black dots indicate positions of the roots while the coloured curves (blue, green and orange) show the trace of the tips of cilia. Blue and green traces indicate the small tilt in cilia (middle right) while the orange trace corresponds to the extensive tilt (bottom right).
Reprinted with permission from Ref.~\cite{Nonaka2005}, licensed under \href{https://creativecommons.org/licenses/by/4.0/}{CC BY 4.0}.} \label{nodal}
\end{figure}

   { In this subsection, we first} focus on the specific case of  nodal cilia. These are found at the ventral surface of the node during the early stage of the embryo development. Nodal cilia are believed to cause the breaking of the left-right body symmetry by creating a leftward flow of the extra-embryonic fluid \cite{LRsymm}. The internal architecture of nodal cilia is slightly different than that of primary cilia.{ There is no central pair of microtubules,  which makes nodal cilia beat in such a way that they trace a tilted cone-like surface. }
  The tilt of the orbit can be quite pronounced, as illustrated in Fig.~\ref{nodal} from mice experiments in Ref.~\cite{Nonaka2005}. The tilt in the mice embryos was measured to be $\theta = 40^\circ \pm 10^\circ$  in 
 Ref.~\cite{Buceta2005}, while   values in in the range  $\theta\in [15^\circ, 35^\circ]$  have also been reported in the literature~\cite{Nonaka2005}. Changing the mean tilt turns out to not affect  our final results considerably, so, to capture the effect of the tilt variability on the synchronisation, we assume  $\theta_1 = 40^\circ$ and vary the value of  $\theta_2$ between $10^\circ$ and $70^\circ$.{  It should be noted that our far-field model cannot describe the cases of nodal cilia that physically touch the epithelium cells (see the extensive tilt in Fig.~\ref{nodal}) and thus the results that follow would not be applicable to such cases.}

We consider two specific limits in which the results are compared. In one, we  assume that the vertical extend of the orbit is much smaller than its height above the lower wall ($R \sin \theta_i \ll h$) while in the other, more realistic limit, we  assume that those quantities are  of similar magnitude,  i.e.,~$R \sin \theta_i \lesssim h$.{ Note that, in the first case, the range of the tilt angle $\theta_i$ is still assumed to be that of nodal cilia (quoted above). This   is  meant to represent the small radius $R$ limit, a limit that is not particularly biologically motivated but does stress the difference in the synchronisation mechanisms, as it is shown below. The results of the elastic compliance mechanism are independent of the choice of the limit, but it makes a substantial difference in the results for the force modulation mechanism (see calculations in Sec.~\ref{FM}).}

Under the assumption $R\sin\theta_i\ll h$,  we obtain that the force modulation mechanism leads to the same state of synchronisation, with or without the upper wall, for any values of $\theta_1, \theta_2$, and $\alpha$. This can be seen from Eqs.~\eqref{FMpotentials} since both potentials have minimum at either $0$ or $\pi$, with a stable point that  depends only the sign of $\sin \Delta_0^{(1)}$ and $\sin \Delta_0^{(2)}$. From Eqs.~\eqref{sin2} and \eqref{sin1}, it is straightforward to see that these always have the same sign as
\begin{equation}
	\sin \Delta_0^{(1)} \sin \Delta_0^{(2)} = \frac{1}{2}r_1^{-1} r_2^{-1} (\cos \theta_2-\cos \theta_1)^2\sin^2 2\alpha\geq 0.
\end{equation}

\begin{figure}[t]
	\centering
\includegraphics[width=\linewidth]{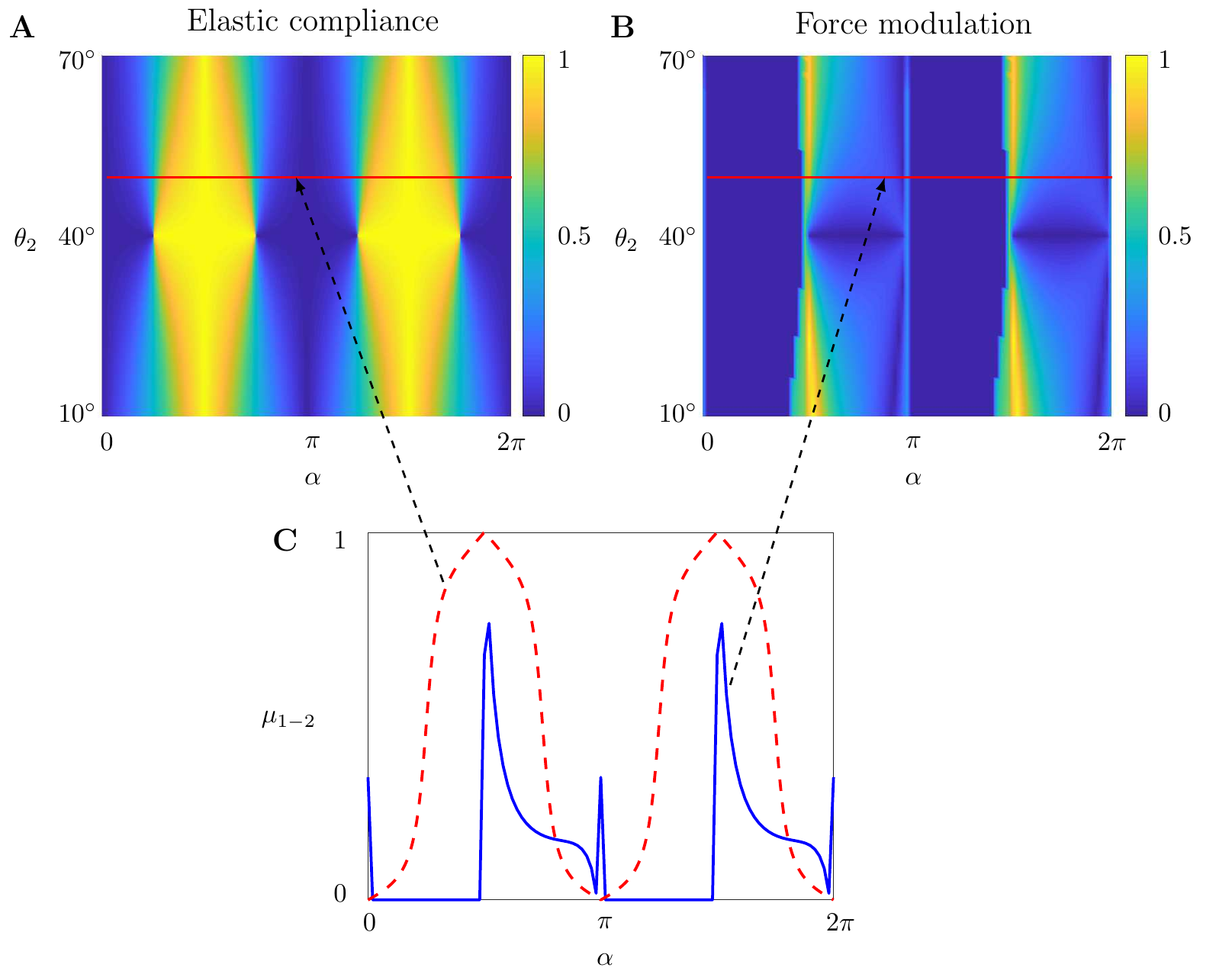}
	\caption{Iso-values of the normalised phase difference,
	$\mu_{1-2}$,  comparing the stable phases to which the oscillators synchronise in cases of a single wall and two walls for the elastic compliance ({\bf A}) and force modulation ({\bf B}) mechanisms. Colours indicate the value of $\mu_{1-2}$ in the case $\alpha_1=\alpha_2=\alpha$ and $\theta_1=40^\circ$. The red line shows the location of the $\theta_2=50^\circ$ section   shown in ({\bf C}).}\label{comparisonfig}
\end{figure}

 Consequently, the force modulation mechanism is absolutely robust to confinement and the synchronisation dynamics is unchanged by the presence of the second nearby surface. In contrast, the elastic compliance mechanism does not provide the same level of robustness, as illustrated in Fig.~\ref{comparisonfig}~{\bf A} for fixed $\theta_1=40^\circ$. There, we see a plot of the measure $\mu_{1-2}$, introduced above, for the elastic compliance mechanism. We fix the orbit of the first sphere to have the tilt $\theta_1=40^\circ$ equal to the average tilt of nodal cilia, and let the tilt of the second one vary between $10^\circ$ and $70^\circ$. We also vary the angle $\alpha=\alpha_1=\alpha_2$, which corresponds to the angle between the pumping direction and the vector connecting centres of the orbits. Warmer colour in the plot indicates a higher value of $\mu_{1-2}$ and thus a lesser level of robustness. Visually, we can notice that for about a half of the $\{\alpha,\theta_2\}$ parameter space the elastic compliance mechanism leads to a similar synchronisation state in both cases of confinement while for the other half of the space it leads to very different states. We will quantify this observation at the end of this section.

Under the different geometrical limit $R\sin \theta_i \lesssim h$, the coefficients in the potential from Eq. \eqref{Potential2} become
\begin{eqnarray}
C_0^{(2)} &=&  -\frac{1}{2}\cos 2\alpha
(\sin \theta_1-\sin \theta_2), \\
B_1^{(2)} &=& -\frac{1}{8} \sin 2\alpha [-6 \sin
({\theta_1}+{\theta_2})+\sin 2 {\theta_1}+\sin 2 {\theta_2}], \\
B_2^{(2)} &=& -\frac{1}{4} \sin 2\alpha [1-\cos(\theta_1-\theta_2)] \sin
({\theta_1}+{\theta_2}), \\
A_1^{(2)} &=& -\frac{1}{4} (\sin \theta_1-\sin \theta_2)\cos 2\alpha (\cos
\theta_1 \cos \theta_2-3), \\
A_2^{(2)} &=& -\frac{1}{4} \cos 2\alpha (\sin
\theta_1-\sin \theta_2) (1-\cos \theta_1 \cos \theta_2),
\end{eqnarray}
where the superscript $(2)$ is indicates the two-wall case. In case of a single wall, these coefficients reduce to
\begin{eqnarray}
C_0^{(1)} &=&  -\frac{1}{4} (1+\cos 2 \alpha) (\sin \theta_1-\sin \theta_2), \\
B_1^{(1)} &=& -\frac{1}{16} \sin 2\alpha [-6 \sin (\theta_1+\theta_2)+\sin 2\theta_2 +\sin 2\theta_1], \\
B_2^{(1)} &=& -\frac{1}{8} \sin 2\alpha (1-\cos (\theta_1-\theta_2)) \sin (\theta_1+\theta_2), \\
A_1^{(1)} &=& -\frac{1}{8} (\sin \theta_1-\sin \theta_2) \big[\cos 2\alpha(\cos
\theta_1 \cos \theta_2 -3) -3-\cos \theta_1 \cos \theta_2 \big], \\
A_2^{(1)} &=& -\frac{1}{8} (\sin \theta_1-\sin
\theta_2) [\cos 2 \alpha(1-\cos \theta_1 \cos \theta_2) +1+\cos \theta_1 \cos \theta_2].
\end{eqnarray}
Comparing the two results, we see that the normalised potentials satisfy
\begin{equation}
	\hat{V}^{(1)} = \hat{V}^{(2)}-\frac{1}{8}(\sin \theta_1-\sin \theta_2)\left(2 \Delta - (3+\cos \theta_1 \cos \theta_2) \sin \Delta + \frac{1}{2}(1+\cos \theta_1 \cos \theta_2) \sin 2 \Delta \right).
\end{equation}

Bearing in mind that  $\phi_1-\phi_2 = \Phi_1-\Phi_2+O(A)$, we plot  
 in Fig.~\ref{comparisonfig}~{\bf B} the values of the normalised difference in stable phases, $\mu_{1-2}$, for the same choice and range of $\theta_i$ and $\alpha_i$ as was previously set for Fig.~\ref{comparisonfig}~{\bf A}. As detailed above, results for the elastic compliance mechanism are independent of the choice of the geometric limit and hence  Fig.~\ref{comparisonfig}~{\bf A} captures the robustness of this mechanism in the limit currently considered, as well.
  We first note that   dependence on $\theta$  is much weaker  than that on $\alpha$.   Secondly, it  is clear from the figure that the force modulation mechanism (Fig.~\ref{comparisonfig}~{\bf B}) is significantly more robust to confinement than the elastic compliance mechanism (Fig.~\ref{comparisonfig}~{\bf A}). 
 This is further  illustrated in Fig.~\ref{comparisonfig}~{\bf C} where we plot  a section of the graphs for the specific  value $\theta_2=50^\circ$.

To draw a quantitative conclusion on the  robustness of the two mechanisms, we next introduce an integrated measure $\mu_I$ by relying on two  assumptions drawn from   known information on nodal cilia. First, we assume that the roots of cilia are isotropically distributed, i.e., ~that $\alpha$ is uniformly distributed in  the interval $[0,2\pi[$. Secondly, we choose the tilt $\theta_2$ to be normally distributed with mean $40^\circ$ and standard deviation $10^\circ$, thus reflecting the experimental measurements from Ref.~\cite{Buceta2005}. As a result, the integrated measure $\mu_I$ is defined as the integral
 \begin{equation}
    \mu_I = \frac{1}{2\pi C}\int_{0}^{2\pi} d\alpha\int_{10^\circ}^{70^\circ} d\theta_2 \, f_n(\theta_2)\, \mu_{1-2}(\alpha,\theta_2)
 \end{equation}
 where $f_n(x) = (2\pi \sigma^2)^{1/2} \exp [-(x-m)^2/2 \sigma^2]$, with $m = 40^\circ$, $\sigma = 10^\circ$, and $C = \int_{10^\circ}^{70^\circ} f_n(\theta_2)\,d\theta_2\approx 0.9954$. 
 
 Using this definition, we can compute the   values of $    \mu_I $ for the two synchronisation mechanisms. We obtain that   for   elastic compliance   the value is $\mu_I^{EC} = 0.4871$, 
 supporting the visual observation   that for about a half of the parameter space the elastic compliance mechanism leads to a similar synchronisation state in both cases of confinement  ($\mu_{1-2}$ close to 0) while for the other half it leads to very different states ($\mu_{1-2}$ close to 1). In contrast, for the  force modulation mechanism we obtain the much smaller value of $\mu_I^{FM} = 0.1178$, indicating a much higher level of robustness to confinement when compared to elastic compliance.
 {
 	
  \subsection{Primary cilia}

In this second subsection, we investigate the robustness of the two synchronisation mechanisms in the case of primary cilia. Unlike   nodal cilia, primary cilia   have a central pair of microtubules and additional internal structures that break the axial symmetry and guide these active filaments to beat in a plane perpendicular to the substrate to which they are attached. In terms of our model, this translates to assuming $\theta_1 = \theta_2 \approx \pi/2$, i.e.,~taking vertical orbits. We still keep the other assumptions outlined above and cilia beat in the same direction $\alpha_1 = \alpha_1 = \alpha$ with a  maximal drag  obtained at the highest point of the orbit, i.e.,~$\delta = 0$. Under these assumptions the equations for the evolution of the  phase difference  become  simplified and are given in 
Table~\ref{tab::eqs}.

\begin{table}[h]
	\begin{tabular}{l|c|c}
		 & One wall & Two walls \\ \hline
		 EC $\dot{\Delta}/\tau \omega^2 = $ & $-\tilde{\beta}\cos^2 \alpha \sin \Delta $ & $-\beta \langle y_1,y_2 \rangle \cos 2\alpha \sin \Delta$\\
		 FM $\langle\dot{\Delta}\rangle / \Omega A^2 = $~ & $-\frac{hR}{2H^2} \cos^2 \alpha (1+2\cos \Delta) \sin \Delta $ &  $-\frac{R^3(2h-H)}{2H^4} \cos 2\alpha (\cos^2 \Delta +2B\cos \Delta +B) \sin \Delta$ \\
	\end{tabular}
	\caption{{Equations for the evolution of the phase difference  for different levels of confinement (one vs two walls) and different synchronisation mechanisms, i.e.,~elastic compliance (EC) vs force modulation mechanism (FM). Note that $B= (1+2(h-H)h R^{-2})/2$.}} \label{tab::eqs}
\end{table}

We first  note that both mechanisms in either case of confinement do exhibit stable fixed points of the phase difference, meaning that in all cases the system synchronises to a fixed phase-difference. For a single wall, both mechanisms
lead to the in-phase synchronisation for almost all pumping directions $\alpha$ (excluding  when $\cos \alpha = 0$, which is a set of measure zero). On the other hand, in the case of two walls,  the value of the ultimate phase difference depends not only on the mechanism but also on the sign of $\cos 2\alpha$. Specifically, the elastic compliance mechanism with two walls leads to in-phase synchronisation if $\cos 2\alpha>0$ but   opposite-phase    ($\Delta = \pi$)  if $\cos 2\alpha<0$. The force modulation mechanism similarly leads to in-phase synchronisation if $\cos 2\alpha>0$ but if $\cos 2\alpha<0$ the system synchronises to a  phase difference whose value depends on the geometry of the orbit; it is given by $\Delta = \pm \Delta_0 (h,R)$ with
\begin{equation}
	\cos \Delta_0 = -\frac{1}{2} +\frac{h(H-h)}{R^2} - \left(\frac{h^2(H-h)^2}{R^4} -\frac{1}{4} \right)^{1/2}.
\end{equation}
 Interestingly, having in mind the geometrical limits  $R<h$ and $R<H-h$ (i.e.~the orbit has to fit between the walls), the range of possible values for this geometry dependent phase difference is relatively narrow, $1.94 < \Delta_0<2\pi/3 \approx 2.09$ (in radians), and thus the dependence on the non-orientational parameters of the orbit is weak.
 
Even though the mathematical results in the case of primary cilia are  clearer than for the nodal cilia, the results for the robustness to  synchronisation are not as conclusive. In the geometric case $\cos 2\alpha>0$, when cilia are pumping in a direction relatively close to the one joining their anchoring points (see illustrative summary  in Fig.~\ref{fig::primarycilia}), both  mechanisms  lead to in-phase synchronisation in   both levels of confinement; we  thus have perfect robustness to confinement in that case. In contrast, when $\cos 2\alpha < 0$,  the elastic compliance mechanism leads to   opposite-phase synchronisation in the strongly confined setup so the robustness measure (see the start of the section) is $\mu_{1-2}=1$. In the same situation, force modulation leads to a phase-locked state with a phase difference $\Delta_0\approx 2\pi/3$ and thus a smaller robustness measure of $\mu_{1-2}\approx2/3$. Unlike the nodal cilia case, it is therefore more difficult to definitely gauge the dominance of one mechanism over the other. Notably, this  difference between the phase-locked states of the two cilia  ($\pi$ and $\Delta_0\approx2\pi/3$)  obtained in half the   parameter space (the domain where $\cos 2\alpha<0$) might be important in the collective organisation of a large ciliary array.   
\begin{figure}
	\includegraphics[width=\linewidth]{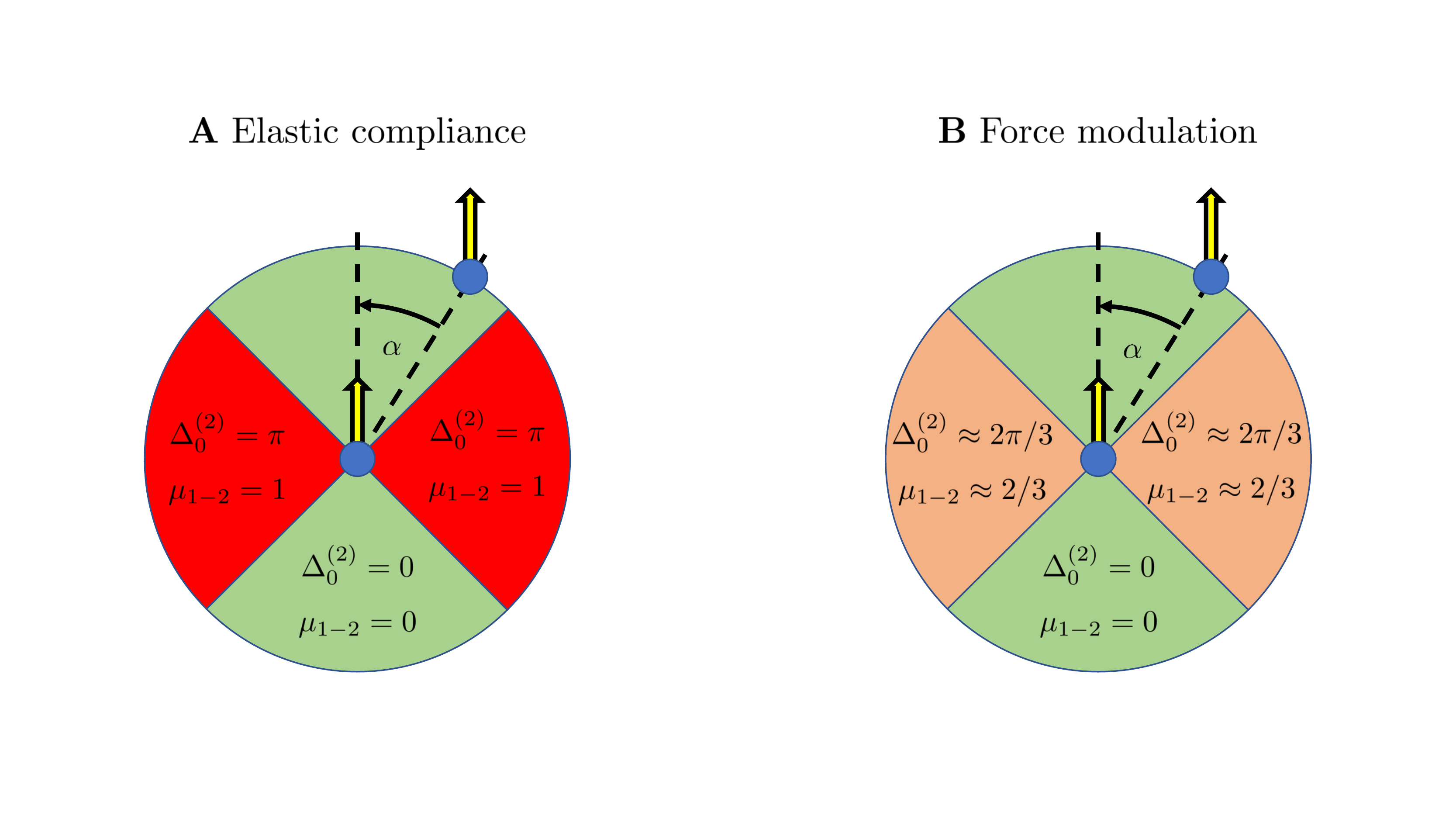}
	\caption{ Illustration of the dependence of the stable phase difference under strong confinement ($\Delta_0^{(2)}$) and the robustness measure ($\mu_{1-2}$) on the relative position of two primary cilia. Results for the elastic compliance and force modulation mechanisms are shown separately in A and B, respectively.}\label{fig::primarycilia}
\end{figure}
}
\section{Discussion\label{dicussion}}

It is now well accepted that the   synchronisation of cilia is due to a combination of  long-range hydrodynamic interactions with  physical mechanisms allowing the ciliary phases  to evolve. Two of such mechanisms have been identified by previous work, the elastic compliance of the periodic orbit or oscillations driven by phase-dependent biological forcing, both of which can lead generically to stable phase-locking. In this paper, we used minimal models of cilia to theoretically investigate  the effect of  strong  confinement on the effectiveness of   hydrodynamic synchronisation. 

Specifically we have compared the usual ciliary dynamics near a single no-slip wall to the dynamics arising when a second nearby surface  is introduced. We called this confinement strong because the distance between the surfaces is much smaller than the typical distance between the cilia.  We computed  separately the impact of hydrodynamic confinement on  the synchronisation dynamics of  the elastic compliance and the force modulation   mechanisms and compared our results to the standard case where the cilia are orbiting near a single surface. {Applying first our results to the biologically-relevant situation of nodal cilia rotating near surfaces, we show that force modulation is a mechanism with a   higher chance, than the elastic compliance mechanism, to lead to  similar phase-locked states under strong confinement to those without confinement. Our results point therefore to the robustness of force modulation for synchronisation of nodal cilia, an important feature for biological dynamics suggesting that it could be the  physical mechanism most essential overall in these ciliary arrays. In the second biologically relevant situation of primary cilia, whose beat patterns are qualitatively different from those of nodal cilia, results on robustness are not as pronounced but still  favour   the force modulation  mechanism. }

Although, to the best of our knowledge,  experiments have   yet     to  investigate  issues of synchronisation under strong  confinement, recent results using the{ algae of the genus} {\it Chlamydomonas} confined in channels about twice their sizes  have shown the important role  of  the 
higher moments of the distribution of viscous forces in the far-field flow of these motile cells~\cite{PolinTC}. Combined with these results,  the modelling approach presented in our paper could  be  adapted to study flagellar  synchronisation in the case where the  cells on which the flagella are anchored  are free  to move in  the fluid.

\begin{acknowledgments}

This project has received funding from the European Research Council under the European Union's Horizon 2020 Research and Innovation Programme (Grant No.~682754 to E.L.) and from  Trinity College, Cambridge (IGS scholarship to I.T.).
	
\end{acknowledgments}
\newpage

\end{document}